\documentclass[%
preprint,
preprintnumbers,
amsmath,amssymb,
aps, showkeys
]{revtex4-2}

\usepackage{graphicx}
\usepackage{dcolumn}
\usepackage{bm}
\usepackage[colorlinks=true, citecolor=blue, linkcolor=blue, urlcolor=blue]{hyperref}

\usepackage[separate-uncertainty=true]{siunitx}
\DeclareSIUnit\angstrom{\textup{~\AA}}

\usepackage{overpic}

\usepackage{upgreek}
\usepackage{bm}
\usepackage{nicefrac}
\usepackage{calligra}
\usepackage{xspace}

\newcommand{\xray}{X-ray\xspace}
\newcommand{\xrays}{X-rays\xspace}
\newcommand{\ie}{\textit{i.\,e.}\xspace}
\newcommand{\eg}{\textit{e.\,g.}\xspace}
\newcommand{\cf}{\textit{cf.}\xspace}
\renewcommand{\P}[1]{\mathcal{P}^{#1}}
\newcommand{\A}[1]{\mathcal{A}^{#1}}
\newcommand{\isoG}[2]{\mathcal{G}(#1, #2)}
\newcommand{\isoF}[2]{\mathcal{F}(#1, #2)}
\newcommand{\isog}{\mathcal{G}}
\newcommand{\isof}{\mathcal{F}}

\begin{document}

\title{Linearization Routines for the Parameter Space Concept to determine Crystal
Structures without Fourier Inversion\\(Centrosymmetric cases in two and three-dimensional parameter space)}

\author{Muthu Vallinayagam}
\affiliation{Technical Physics, University of Applied Sciences, Friedrich-List-Platz 1, D-01069 Dresden, Germany\\
Center for Efficient High Temperature Processes and Materials Conversion ZeHS, TU Bergakademie Freiberg, Winklerstr.~5, D-09596 Freiberg, Germany }

\author{Melanie Nentwich}
\email{Melanie.Nentwich@desy.de}
\affiliation{Deutsches Elektronen-Synchrotron DESY, Notkestr.~85, D-22607 Hamburg, Germany}

\author{Dirk C. Meyer}
\affiliation{Center for Efficient High Temperature Processes and Materials Conversion ZeHS, TU Bergakademie Freiberg, Winklerstr.~5, D-09596 Freiberg, Germany\\Institute of Experimental Physics, TU Bergakademie Freiberg, Leipziger Str.~23, D-09596 Freiberg, Germany}

\author{Matthias Zschornak}
\email{matthias.zschornak@physik.tu-freiberg.de}
\affiliation{Technical Physics, University of Applied Sciences, Friedrich-List-Platz 1, D-01069 Dresden, Germany\\
Center for Efficient High Temperature Processes and Materials Conversion ZeHS, TU Bergakademie Freiberg, Winklerstr.~5, D-09596 Freiberg, Germany \\ Institute of Experimental Physics, TU Bergakademie Freiberg, Leipziger Str.~23, D-09596 Freiberg, Germany}

\collaboration{Dedicated to our revered teacher Prof. Karl Fischer}

\date{\today}

\begin{abstract}
We present detailed elaboration and first generally applicable linearization routines of the \textit{Parameter Space Concept} (PSC) for determining 1-dimensionally projected structures of $m$ independent scatterers. This crystal determination approach does not rely on Fourier inversion but rather considers all structure parameter combinations consistent with available diffraction data in a parameter space of dimension $m$. The method utilizes $m$ structure factor amplitudes or intensities represented by piece-wise analytic hyper-surfaces, to define the acceptable parameter regions. By employing the isosurfaces, the coordinates of the point scatterers are obtained through the intersection of multiple isosurfaces. This approach allows for the detection of all possible solutions for the given structure factor amplitudes in a single derivation. Taking the resonant contrast into account, the spatial resolution achieved by the presented method may exceed that of traditional Fourier inversion, and the algorithms can be significantly optimized by exploiting the symmetry properties of the isosurfaces. The applied 1-dimensional projection demonstrates the efficiency of the PSC linearization approach based on fewer reflections than Fourier sums. The Monte-Carlo simulations, using the projections of various random two- and three-atom structure examples, are presented to illustrate the universal applicability of the proposed method. Furthermore, ongoing efforts aim to enhance the efficiency of data handling and to overcome current constraints, promising further advancements in the capabilities and accuracy of the PSC framework.
\end{abstract}

\keywords{Crystal structure determination, Two- and three-dimensional parameter space, Centrosymmetric structure, Linearization, Parameter Space Concept, Monte-Carlo simulation, One-dimensional projection, X-ray diffraction, Python}

\maketitle

\section{Introduction}

Solving crystal structures from diffraction intensities plays a vital role in many areas of solid-state research, \eg physics, chemistry, mineralogy, materials sciences, biology, and pharmacy, as it forms the fundamental basis for understanding the properties of materials as well as their effects and functionalities. The corresponding databases grow by tens of thousands of structures every year. The state-of-the-art structure determination methodology is based on \textit{Fourier Inversion} (FI) of the scattering density (\eg electron density for \xray diffraction, nuclear density for neutron diffraction). In the early days, the developments in crystallography were mainly based on computationally efficient FI techniques either directly or indirectly such as the charge flipping method~\cite{Oszlanyi2004}, algebraic method~\cite{Rothbauer1994, Rothbauer1995, Rothbauer1998}, geometrical methods~\cite{Navaza1979}, analytical function methods~\cite{Cervellino2005}, fit methods such as Rietveld refinement~\cite{Tobyactacrys}, and matching learning algorithms~\cite{shiactacrys, Munteanuactacrys, Billing}. 

The currently used FI techniques to construct electron density systematically introduce noise and errors in the calculation due to series termination. Therefore, a large number of terms in the FI series are required, which in turn necessitates a substantial set of experimental observations. Furthermore, the quality of \xray diffraction intensities greatly influences the FI series, with weaker observations contributing less. However, the experimental database is in most cases incomplete since only the quadratic amplitudes of the Fourier coefficients (\ie the structure factors via the reflection intensities) can be determined, and the well-known phase problem of crystallography makes the structure solving more challenging~\cite{Harrison1993, Fischer2005, Fischer2008, Fischer2009}.

\begin{figure}
    \centering
    \includegraphics[width=0.8\textwidth]{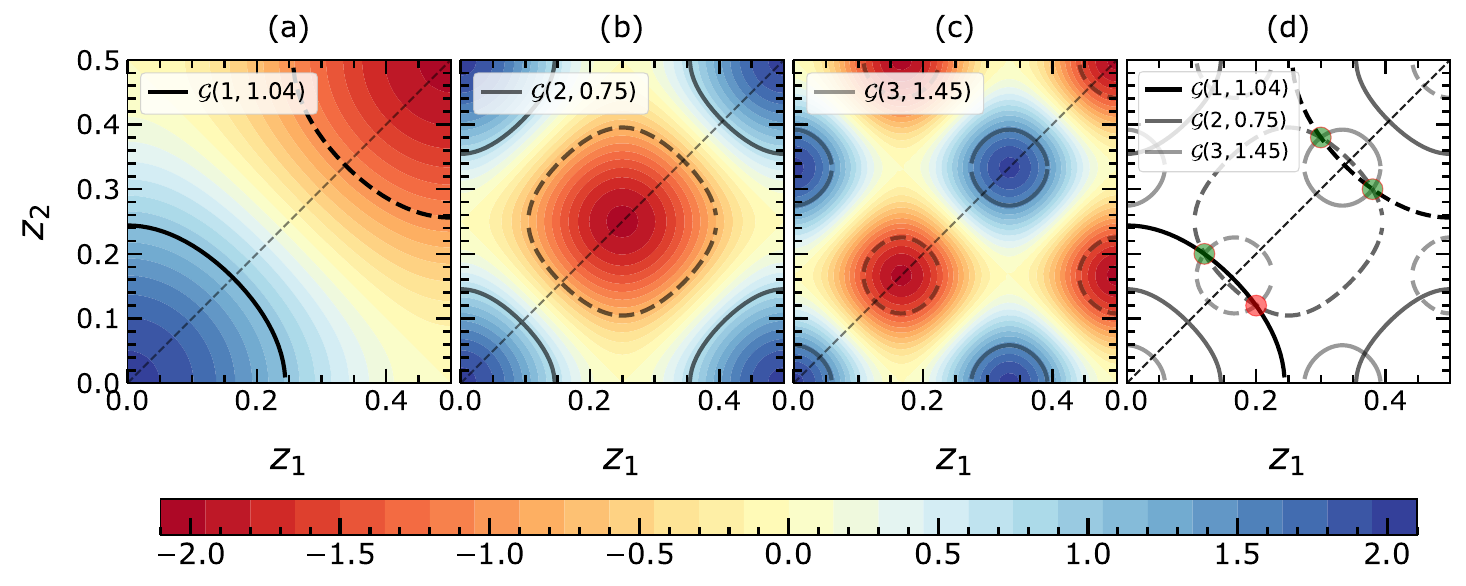}
    \caption{Basic explanation of the \textit{Parameter Space Concept} (PSC). (a)--(c)~2-dimensional Parameter Space for the projection of a crystal structure of two equal scatterers for the projection onto the $z$~axis for reflections $00l, l = 1, 2, 3$. The color map represents the calculated amplitude for each combination of atomic coordinates. The isosurfaces for positive and negative amplitudes generated by the arbitrary atomic coordinates $[0.2, 0.12]$ are highlighted with solid and dashed lines, respectively. (d)~Overlay of the isosurfaces from (a)--(c) with their intersecting points highlighted by red and green points. The intersection point at the red circle is the intended structure to be found, and alternative solutions are at the green circles.}
    \label{fgr:isoex1}
\end{figure}

In order to overcome the demerits of FI techniques, alternative methods have been developed. In this study, we examine the relationship between the structure factor and the atomic positions in crystal structures under the aspect of geometrical correlations. In general, an $m$~atomic structure has $3m$~free positional parameters to be determined, which includes the $x$, $y$, and $z$~coordinates of all $m$~atoms. To simplify the structure-solving process with PSC, the task is split into several independent 1-dimensional projections (in real space), each providing the solution of $m$~parameters within an $m$-dimensional \textit{Parameter Space} (PS; space of atomic coordinates with the orthogonal basis in $\mathbb{R^{\textit{m}}}$)~\cite{Zschornak2023, Fischer2005, knop_thesis, pliz_thesis, Ott1927}. Each point in the PS corresponds to a possible combination of atomic coordinates (\eg projected onto the $z$~axis) and generates a unique \xray diffraction intensity for a predefined reflection. Vice versa, the set of points that reproduce the experimentally observed intensity of a particular reflection defines a manifold called \textit{isosurface}, see Fig.~\ref{fgr:isoex1}(a)--(c). The intersection point from all isosurfaces of different reflections expresses the intended structure, see Fig.~\ref{fgr:isoex1}(d).

The theory of PSC has been developed during the past two decades, mainly focusing on equal atoms~\cite{knop_thesis, pliz_thesis, Ott1927} and aiming to achieve higher spatial resolution than the FI techniques while at the same time using a reduced number of available intensity data sets. Apart from these advantages, PSC will always recognize all possible solutions, which can reproduce the given experimental intensities, although at the cost of parameterizing functions in continuous high-dimensional spaces. Hence, the PSC provides an elegant but computationally expensive method to solve the crystal structures in a step-wise approach splitting the full structure into 1-dimensional projections.

In the present work, we develop further theoretical approaches to treat all aspects of $m$-dimensional PSs, in particular generally applicable linearization routines to parameterize the isosurfaces for efficient functional handling as well as computational storage and determination of intersections. The PSC algorithms are now enhanced to treat the artificial values of the atomic scattering factor of scatterers in the $m$-dimensional PS. The implemented capability to overcome the previously employed \textit{Equal Point Atom} (EPA) model (see Sec.~\ref{sec:strcturefactor}) improves the PSC towards a generally applicable structure determination approach. However, to test the developed algorithms and code, in this manuscript, we determine the centrosymmetric structures so far only in 2- and 3-dimensional PS. Nevertheless, on the first estimation, the derived equations may be straightforwardly enhanced towards higher dimensions, which will be the focus of our continuous research efforts.

This study is structured as follows: the subsequent sections provide an in-depth exploration of our approach, commencing with the fundamental theory underlying the PSC-based framework. Next, the steps involved in solving a maximum of $3$~parameters in the 1-dimensional projection are depicted, including the general description of the linear approximation of isosurfaces. Finally, we present noteworthy generalizations based on Monte-Carlo simulations for the structure determination within the 2- and 3-dimensional centrosymmetric PS.

\section{Theoretical methods}\label{sec:theory}

Solving the crystal structure is a meticulous process that involves the precise determination of the position of each atom within the crystal. In the context of PSC analysis, an effective strategy is employed, making use of a linear approximation of the isosurfaces defined by intensities derived from \xray diffraction data within the corresponding parameter space. The careful analysis of the experimentally measured intensity of each reflection facilitates the reconstruction of the potential combinations of atomic coordinates by utilizing sophisticated piece-wise analytic hyper-surfaces~\cite{Fischer2005, Zschornak2023}.

\subsection{Parameter space and isosurfaces}\label{sec:PS_isosurfaces}

The complete solution of a structure with $m$~atoms in the unit cell consists of $3m$~parameters; accounting for the three distinct coordinate components ($x$, $y$, and~$z$) of each atom. The $3m$-dimensional space containing all possible combinations of those parameters is called the \textit{Parameter Space}~(PS). By employing a 1-dimensional projection of the crystal structure onto the main axes, the complex task of solving $3m$~coordinates is separated into several distinct steps, each solving $m$~coordinates in the $m$-dimensional PS~$\P{m}$. Then assuming oblique reflections (more than one non-zero component), the relationship between the $x$, $y$, and $z$ components of atomic coordinates can be assigned~\cite{Fischer2008}. However, the necessary steps to combine the projections into a complete structure solution will be the topic of a further manuscript. 

Here, we adopt the projection onto the $z$~axis as a representative projection for all axes, while employing the Miller index~$l$ to abbreviate the relevant \xray reflections~$00l$. In general, the PS~$\P{m}$ consists of $m$~orthonormal axes, which we assign to the $z$~components~$z_1, \ldots, z_m$ of the atomic coordinates. When considering a specific reflection, each point $\bm{z} = (z_1, \ldots, z_m)$ in $\P{m}$ has a uniquely defined amplitude, which is directly related to the intensity, see Sec.~\ref{sec:strcturefactor}. However, a specific amplitude value can be achieved through several~$\bm{z}$, resulting in a so-called \textit{isosurface}. The isosurface embodies the entirety of possible combinations of atomic coordinates that generate the same amplitude for a specific reflection within the intricate landscape of~$\P{m}$. Each reflection provides insights into the possible configurations of the crystal structure: only the well-defined combination of atomic coordinates described by all the isosurface can create the experimentally determined amplitude. The precise structure solution is, thus, defined as the intersection of the isosurfaces corresponding to different reflections, see Fig.~\ref{fgr:isoex1}(d). Through detailed and precise analysis, we unravel the regions where these isosurfaces intersect, which are ultimately interpreted as the coordinates of the atoms within the crystal lattice. For error-free intensity values and using the isosurfaces directly, only $m$~different reflections are required to solve for $m$~coordinates, realized as the intersection of $m$~isosurfaces. However, in the cases of linearized isosurfaces (see Sec.~\ref{sec:linearization}) as well as experimental and thus naturally erroneous values, the isosurfaces gain volume and, thus, also the intersection region. An exact point solution cannot be determined, however, adding more reflections will improve the accuracy of the solution.

\subsection{The structure factor}\label{sec:strcturefactor}

The basic need for the structure determination is the information about the Miller index and its corresponding intensity~$I$, which can be expressed as $I \propto|F^2|$, with the \textit{Structure Factor}~$F$. Within this manuscript, we will constrain the discussion to centrosymmetric cases, which provides the advantage that the solution is not a complex value, but lies in the real number space. Thus, we can express the structure factor of the crystal structure as~\cite{Fischer2005, Fischer2006}:

\begin{equation}
    F(l) = 2 \cdot \sum_{i=1}^{m} f_i(l) \cos(2 \uppi l z_i) = 2 \cdot s(l) \cdot \left| \sum_{i=1}^{m} f_i(l) \cos(2 \uppi l z_i) \right|, \label{eq:F}
\end{equation}

where $l$ is the Miller index of reflection~$00l$, $m$~is the number of atoms in the unitcell, $s(l)$~is the sign of the expression, $f_i$~is the atomic structure factor (a real number, considering only Thomson scattering without resonance corrections), and $z_i$ is the coordinate of the atom with index~$i$. 

If all atoms are considered to be equal (and point-like), then their scattering factors are identical $f_i(l) = f(l)$ and can be set to unity; hence, Eq.~\eqref{eq:F} becomes 

\begin{align}
    G(l) &= 2 \cdot s(l) \cdot \left| \sum_{i=1}^{m} \cos(2 \uppi l z_i) \right| = 2 \cdot s(l) \cdot g(l), 
\end{align}

where

\begin{align}
    g(l) = \left| \sum_{i=1}^{m} \cos(2 \uppi l z_i) \right| \label{eq:g}
\end{align}

is the modulus of the \textit{Geometric Structure Factor}~$G$. Detailed relationships for atoms on special positions are discussed by Fischer \textit{et al.}~\cite{Fischer2005}. The latter case of identical point-like atoms is referred to as the \textit{Equal Point Atom} (EPA) model hereafter. The case of realistic scattering factors will be referred to as the non-EPA model. 

It is noteworthy to mention the difference between the isosurface, \ie the manifold, and the structure factor in terms of parameter dependencies. The \textit{isosurface} is a subspace of dimension $m-1$ in $\P{m}$, fulfilling a boundary condition of coordinates $z_i$ with a certain amplitude or intensity. For experimentally observed or theoretically calculated amplitudes $2\,g_0(l)$ and $\left|F_0(l)\right|$, the isosurface of the geometric structure factor is mathematically expressed for the EPA case as:

\begin{align}
    \isoG{z_i, l}{g_0(l)} \text{\quad satisfying \quad } g(l) - g_0(l) = 0 \label{eq:G}
\end{align}

and of the structure factor for the non-EPA case as:

\begin{align}
    \isoF{z_i, l}{\nicefrac{\left|F_0(l)\right|}{2}} \text{\quad satisfying \quad } \left|F(l)\right| - \left|F_0(l)\right| = 0.\label{eq:Fintensity}
\end{align}

For better readability we use the expressions $\isoG{l}{g_0(l)}$ or $\isoF{l}{\nicefrac{\left|F_0(l)\right|}{2}}$ for EPA or non-EPA cases, respectively, where only $l$ and $g_0(l)$ or $\nicefrac{\left|F_0(l)\right|}{2}$ are specified explicitly. Further, the (geometric) structure factors $G(l)$ or $F(l)$ represent fixed structures $z_i$ and only depend on $l$. Otherwise, simply $\isog$ or $\isof$ are used for EPA or non-EPA cases. The isosurface can be defined from given amplitudes~$2 \, g(l)$, $\left|F_0(l)\right|$ or intensities~$\left|2 \, g_0(l)\right|^2$, $\left|F_0(l)\right|^2$, which is marked with the respective index $A$ or $I$ on $\isog$ or $\isof$. 

The value of the amplitude~$g(l)$ for a given reflection~$l$ is visualized in dependence of the atomic coordinates~$z_1$ and~$z_2$ as the color map in Fig.~\ref{fgr:isoex1}. Also, Fig.~\ref{fgr:isoex1} shows an isosurface corresponding to an arbitrary set of amplitudes $g_0(l)$, represented by black lines. To account for the complexities of realistic applications including different atoms and their scattering behavior, we will also consider various combinations of scattering factors. Although the $f_i$'s are complex quantities influenced by factors like the energy of the incoming \xrays and inter-lattice distances \cite{woolfson_hai-fu_1995}, we have carefully chosen appropriate real numbers representing heavy, medium, and light atoms. 

\begin{figure}
    \centering
    \includegraphics[width=0.8\textwidth]{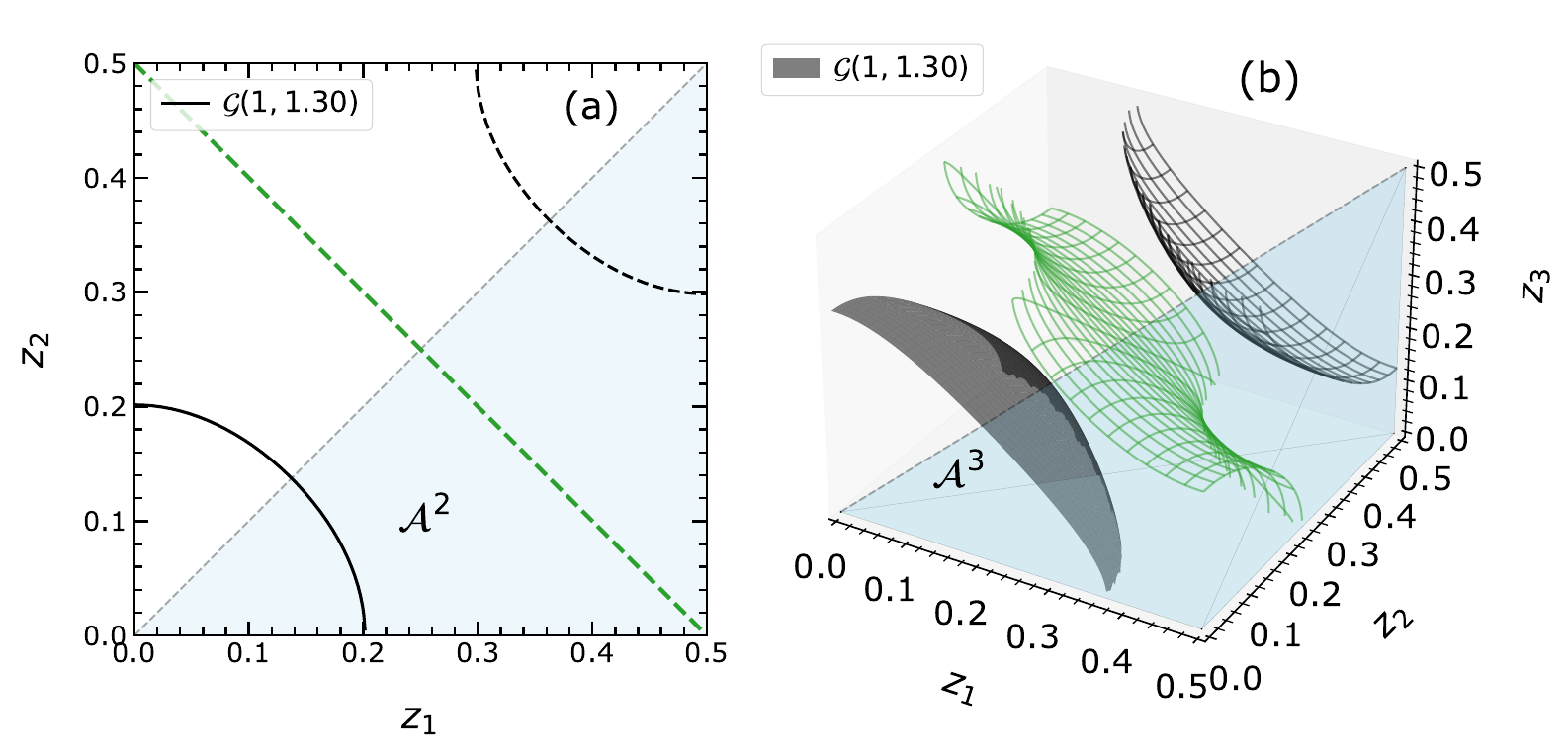}
    \caption{Isosurfaces $\isog$ for reflection $l=1$ in 2- and 3-dimensional PS in (a) and (b) respectively, with highlighted asymmetric parts $\A{2}$ and $\A{3}$. The black continuous line in (a) and solid surface in (b) represent the instance of the isosurface with the positive sign ($s(l) = +1$). The black dotted line in (a) and wireframe in (b) represent the negative instance ($s(l) = -1$). The green shaded regions represent $\A{2}$ and $\A{3}$ of $\P{2}$ and $\P{3}$, respectively and contain the contributions from both $s(l)=\pm 1$. However, the knowledge or selection of $s(l)$ of the amplitude further restricts the $\A{2}$ and $\A{3}$ reducing the volume by a factor of two via the zero $\isoG{l}{g(l)=0}$ isosurface represented by the green dashed line in $\P{2}$ and wireframe in  $\P{3}$.}
    \label{fgr:isoex}
\end{figure}

We realized that varying scattering factors $f_i$ affect the topology of the isosurfaces, \cf Eq.~\eqref{eq:F}. In EPA, $f_i$ are set to unity, hence the isosurfaces~$\isog$ with large $g_0$ values are almost a circle in $\P{2}$ and a sphere in $\P{3}$, \cf Fig.~\ref{fgr:isoex1} and ~\ref{fgr:isoex}. In the non-EPA model, the isosurface~$\isof$ exhibits a variety of topologies, that require careful and independent handling (Sec.~\ref{Topologies of isosurface}). Furthermore, the overall appearance of the isosurfaces in the parameter space is influenced not only by the ratios of scattering factors but also by the reflection index~$l$. As the index~$l$ increases, the number of disjoint isosurface regions also increases, see Fig.~\ref{fgr:isoex1}. Section~\ref{sec:linearization} gives more details on the systematic.

The detailed analysis of intrinsic symmetries can contribute to a further reduction of the computational effort, by reducing the possible solution space~$\P{m}$ to the asymmetric parameter space~$\A{m}$. Those symmetries comprise (1)~the centrosymmetry of the structure, (2)~the permutation symmetry for equal or partially equal atoms, and (3)~the choice of origin. The assumed centrosymmetry leads to the spatial limitation of the PS to the range $[0.0, 0.5]^m \cdot c$, where $c$ is the lattice constant of the crystal in the specified projection. 

\begin{figure}
    \centering
    \includegraphics[width=1\linewidth]{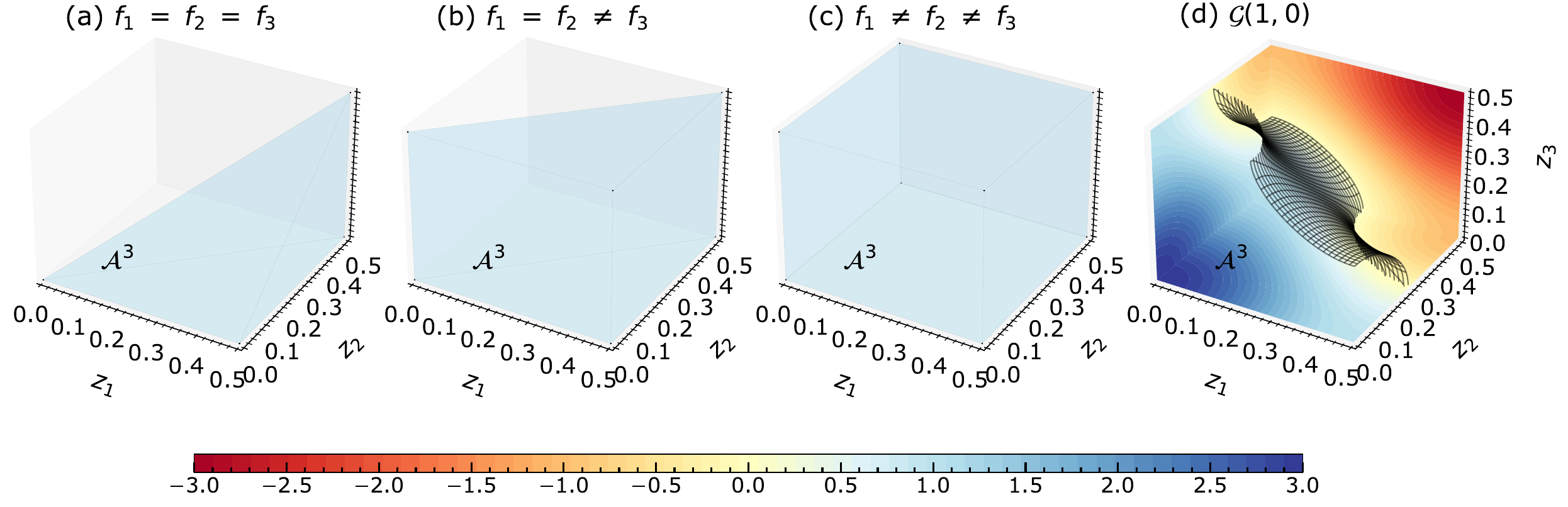}
    \caption{The definition of asymmetric units upon specific sets of $f$ values is shown in (a)--(c). The $\A{3}$ in (a) is defined for the EPA framework where all $f_i$ are equal. The $\A{3}$ in (b) and (c) are defined for the non-EPA framework. In (b) it is assumed that two $f$'s (along $z_1$ and $z_2$ directions) are the same and $f_3$ is different. In (c) it is assumed that all $f$'s are different and hence the $\A{3}$ is equivalent to $\P{3}$. In~(d) the $\isoG{l=1}{g_0(l)}$ are given for the EPA case representing the magnitude of $g_0(l)$ from \numrange{-3}{3} in color code. The \textit{zero isosurface} $\isoG{l=1}{g_0(l)=0}$ separates the PS into two halves containing isosurfaces $\isoG{l=1}{g_0(l)}$ of positive or negative signs, respectively. When applying the choice of origin symmetry, the origin is fixed at $[0,0,0]$. Then the asymmetric unit reduces to half under the zero isosurface, containing only the positive signs $s(l=1)$, for the solution search. The color bar gives the magnitude of $g_0(l=1)$ with the applied sign.}
    \label{fgr:As}
\end{figure}

Furthermore, following the discussion by Fischer~\textit{et al.}~\cite{Fischer2005}, in the case of EPA, the full PS~$\P{m}$ can be reduced by permutation of atomic coordinates, which encompasses both positive and negative instances of the isosurfaces. This reduction can be visualized using the corner points fixed at coordinate $[0,0,\dots,0],~ [0.5,0,\allowbreak \dots,0],~\dots,~[0.5, 0.5, \allowbreak \dots,0.5]$, \cf entire shaded area in Fig.~\ref{fgr:isoex}. Even in the non-EPA cases, the occurrence of $n$~equal atoms will induce limited permutation symmetry within the respective subspace of the PS, for which a reduction of $\P{m-n}$ towards $\A{m-n}$ can be obtained. The possible cases for $m=3$ are shown in Fig.~\ref{fgr:As}. As an example, if $f_1$ and $f_2$ are equal, then the $\P{3}$ is halved and the $\A{3}$ is a triangular prism as shown in Fig.~\ref{fgr:As}(b). 

In addition to the permutation of atomic coordinates, the choice of the origin of the crystal system at one of the two centers of inversion (at $[0,0,\dots,0]$ and $[0.5,0,\dots,0]$) provides another intrinsic symmetry to reduce the possible solution space where the given structure resides. The composition of the set of $f_i$ defines a specific \textit{zero isosurface}, one boundary of the asymmetric unit given by the isosurface $\isoG{1}{g(1)=0}$, \cf Fig.~\ref{fgr:As}(d) as an example for $\P{3}$. Once the choice of origin is utilized only the green region below or above the zero isosurfaces needs to be analyzed. These boundaries simultaneously define the linearization boundaries and the allowed solution space for the structure investigation. 

Additionally, $\P{m}$ can also be reduced using the sign of the isosurfaces if available; represented exemplarily for $l=1$ by the separation of the shaded region in Fig.~\ref{fgr:isoex}(a) and~(b) via the zero isosurfaces $\isoG{l}{g(l)=0}$.

\subsection{General linearization approach within the Parameter Space Concept}\label{sec:steps} 

\begin{figure}[ht]
    \centering
    \includegraphics[width=1\textwidth, trim=0.6cm 0.6cm 0.7cm 0.6cm,  clip]{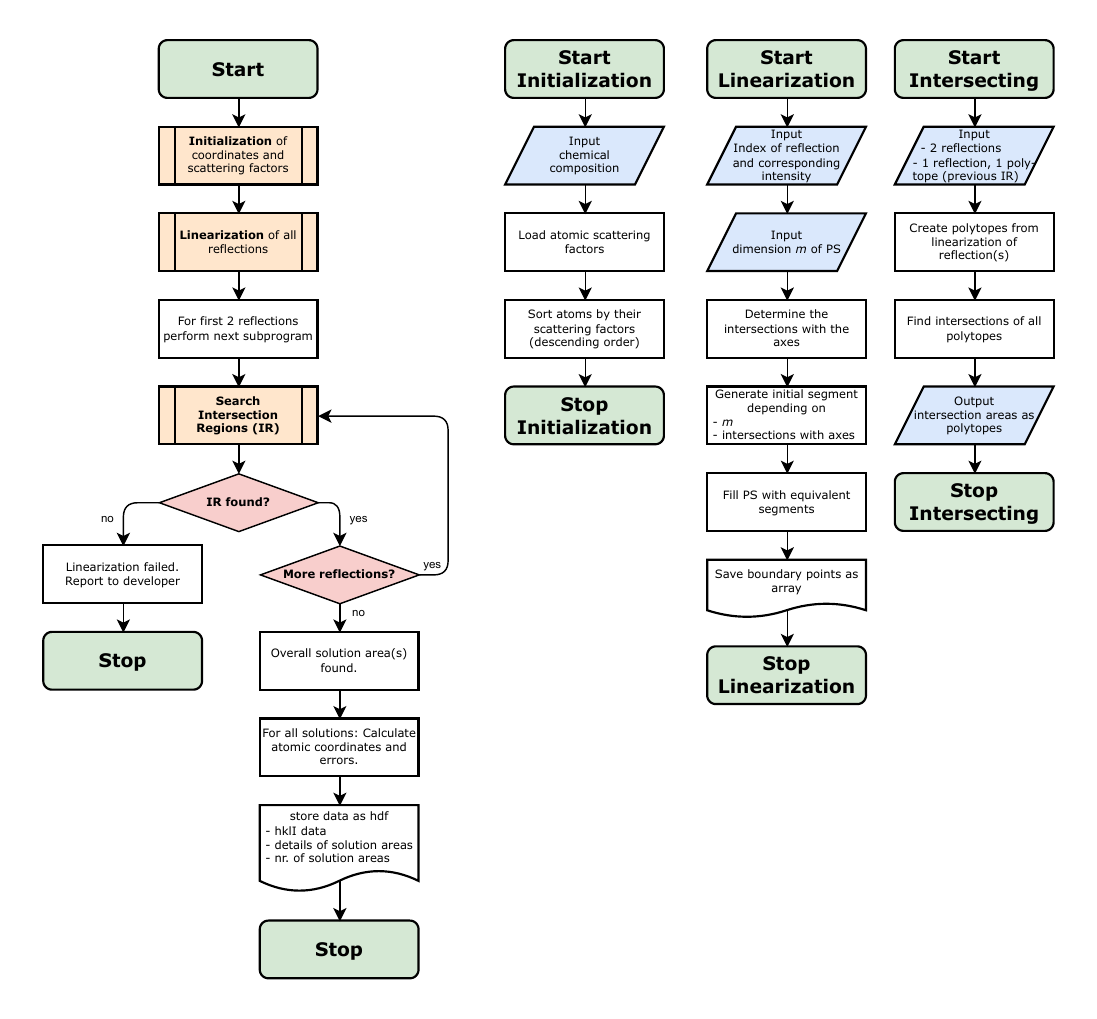}
    \caption{The scheme of operations flow in PSC includes the initialization, linearization, and solution-finding process. Also, many intermediate decisions are made to verify the inclusion of all reflections, reaching intersection regions, imposing an accuracy limit on the area/volume of intersection regions, etc. At the end of the structure determination, detailed outputs including reflections and their polytopes, intermediate intersections results, found solutions, the volume of each polytope, and error on computed $z_i$ are written in HDF-formatted files.}
    \label{fgr:flowchart}
\end{figure}

The primary objective of structure determination in~$\P{m}$ is to identify the intersection of isosurfaces $\isog$ or $\isof$ corresponding to different reflections, see Fig.~\ref{fgr:flowchart}. However, finding the intersection point directly is challenging as it involves complex trigonometric functions, \cf Eq.~\eqref{eq:F}. One possible approach to reduce complexity, which is focused on in this manuscript, consists of employing linearization techniques, that enable the replacement of intricate trigonometric expressions within linear boundaries. This approximation can simplify the problem and utilizes \textit{Set Theory} to explore all potential solution regions. Note that the linear approximation may yield multiple solution regions caused by step-wise intersections of linear approximants of different reflections. The approximation process generally involves the following steps, depicted in Fig.~\ref{fgr:flowchart}.

\subsubsection*{Initialization}

In the first step, we implement specific measures to ensure the flawless execution of the linearization algorithm. This important step involves arranging the atomic structure factors $f_i$ in a descending order, which helps us determine the general curvature of the isosurface~$\isof$, as well as apply the permutation symmetry. The larger~$f_i$ values correspond to heavier atoms and exert maximum control over the behavior of $\isof$, while the smaller $f_i$~values pertain to lighter atoms and thus smaller contributions in the interference dependencies. The influence of~$f_i$ on the curvature of $\isof$ is depicted in Sec.~\ref{Topologies of isosurface}. In this work, we distinguish between different types of topologies: they are called \textit{closed} if $\isof$ intersects all parameter axes and \textit{open} if $\isof$ does not intersect with at least one axis.

\subsubsection*{Linear approximation}

The simplest approximation method is linearization. We bound the curved isosurfaces of each reflection by two straight, parallel lines in~$\P{2}$ or two planes in~$\P{3}$. The normal vector and distance from the origin are the required descriptors of the boundaries and are being determined in this step. The equations and routines for these parameters are described in detail in Sec.~\ref{sec:linearization}. Notably, the linear approximation invokes the \textit{Mean-Value Theorem} at the core~\cite{MVT1, MVT2}. The challenge is the development of a new non-EPA linearization framework, which we present here for the first time giving algorithms to linearize the isosurfaces governed by different atomic scattering factors. This non-EPA framework elevates the capability of PSC to handle the realistic $\xray$ diffraction data. 

\subsubsection*{Solution finding}

Subsequently, the boundary descriptors obtained from the previous step are used to construct polytopes, described by a system of linear inequality equations with the number of variables identical to the dimension~$m$ of the PS. These polytopes are reflecting the PS region that matches the observed intensity for the specific reflection in linear approximation. The goal is to find the common regions that are enclosed by all polytopes of every reflection and that represent the projected atomic coordinates of the structure under investigation. To achieve this final solution, we search for the intersection regions of the polytopes created for successive reflections. In ideal cases, \ie error-free amplitudes for a full set of reflections starting from $l=1$, a single polytope solution region that uniquely represents the given structure may be identified. However, in many cases, the PSC can yield multiple polytope solution regions in a consecutive intersection step for an arbitrary reflection $l$, or when taking into account intensities as observable without the knowledge of the structure factor's phase.

\subsubsection*{Data writing to HDF}

After the solution-finding process, we obtain crucial information such as the volume of solution regions, the coordinates of the edges of these regions, the number of solutions, and the computation time. This data is stored in a file. Additionally, details about processed reflections, the atomic scattering factor, the structure factor, and experimental or theoretical intensity/amplitude are also saved for future reference.\\
We have chosen to use the \textit{Hierarchical Data Format} (HDF) due to its versatility as a data model, making it ideal for managing large and complex datasets~\cite{HDF}. HDF allows for the storage of various types of data within a single file. There is an efficient Python library available that facilitates the integration of this specific file format into our routines. This format guarantees that the data is easily accessible and portable across different platforms, thereby enabling seamless analysis and sharing of results, making it as a superior choice for data storage and retrieval. Furthermore, data stored in HDF files can be easily visualized using the HDFView tool, enhancing the practicality of this choice~\cite{HDF}. More details about HDFView are given in the supplementary material (SM), see Sec.~2 in SM.

\subsection{Linear approximation routines}\label{sec:linearization}

Solving for the intersection point $\bm{z} = (z_1, z_2, \ldots, z_m)$ on a dense grid in  $\P{m}$ will become computationally expensive, challenging, and sometimes even impossible for higher dimensions. However, we can reduce the computational effort significantly by solving for linearized approximations of the isosurfaces. Unfortunately, instead of solution points, linearization introduces expanded solution regions, which ideally should be kept as small as possible. Here, we apply a linear approximation as described by Fischer \textit{et al.}~\cite{Fischer2005}: we develop a basic unit called \textit{segment} to linearize a well-defined part of the isosurface and extend the linearization by shifting and rotating the segment to create a full cover-up of the isosurfaces within~$\P{m}$. The segments consist of a set of parallel boundaries, the inner and outer limits of the isosurface with respect to the PS origin, and limiting boundaries perpendicular to each $z_i$~direction. We have discovered that, in addition to the circle-like topologies considered by Fischer~\textit{et al.} (Fig.~\ref{fgr:isoex}), band-like isosurfaces can also occur. The origin of these and also the different handling within the linearization procedure will be discussed in the following Sec.~\ref{Topologies of isosurface}.

\begin{figure}
    \centering
    \includegraphics[width=1\textwidth]{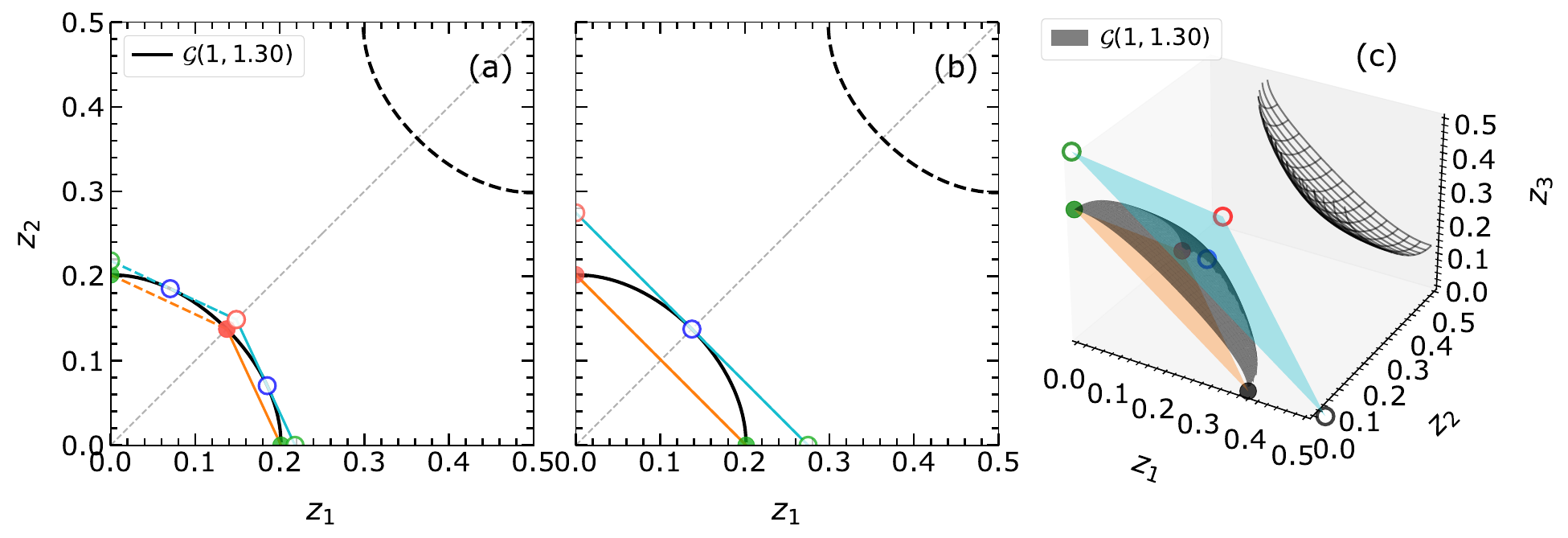}
    \caption{Linearization of an isosurface $\isog$ with (a)~two segments in $\P{2}$, (b)~one segment in $\P{2}$, and (c)~one segment in $\P{3}$. The filled and open circles represent the coordinates of the inner and outer boundary line/plane. The \textit{tangent point} is represented by a blue open circle. The blue dashed line represents the mirror plane in $\P{2}$ along the $[1, 1]$~direction created by the EPA model. An arbitrary value of $1.3$ for the magnitude $g_0$ is selected and the isosurface is shown for reflection $l=1$ in $\P{2}$ and~$\P{3}$.}
    \label{fgr:linearization}
\end{figure}

Examples of the linearization of $\isog$ process for the EPA case are shown in Fig.~\ref{fgr:linearization}. The complete procedure to accomplish linearization is explained in the following paragraphs; they include the generation of segments around the origin of PS as well as the repetition of segments in PS.

\subsubsection{Topologies of isosurface}\label{Topologies of isosurface}

In previous literature, the isosurfaces were always described as circle- or sphere-like, resulting in closed loops in $\P{2}$ and $\P{3}$. Fischer~\textit{et al.} have not introduced the concept of topology as it was not a requirement, since only the EPA models were applied, in which most of the hyper-surfaces are closed circle- or sphere-like. However, during our extensive investigations using Monte-Carlo simulations (see Sec.~\ref{sec:2DMC} and~\ref{sec:3DMC}), we learned that this is not always true, and open, band-like structures can appear.

\begin{figure}
    \centering
    \includegraphics[width=1\textwidth]{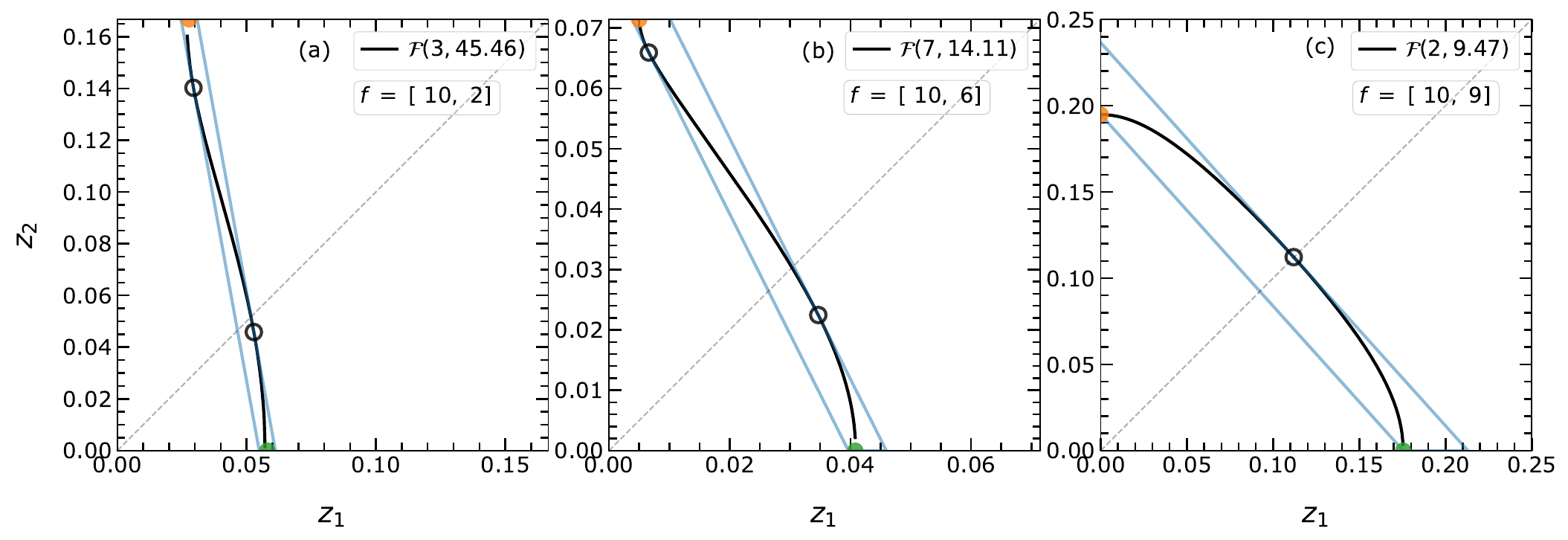}
    \caption{The different possible topologies of isosurface $\isof$ in~$\P{2}$ on varying atomic scattering factors $f_i$ in Eq.~\eqref{eq:F}. The $\isof$ can be categorized according to their intersection along the $z_1$ and $z_2$ directions. The $\isof$ in (a) and~(b) exclusively intersect the $z_1$ axis (open topology) and in~(c) both the $z_1$ and $z_2$ axes at different distances from the origin (closed topology). The developed algorithm handles and approximates these $\isof$'s alike. The boundaries from the approximation are shown by blue lines. The schematic demonstrates the effect of the ratio between $f_i$ on the curvature of $\isof$. The very first segment of the linearization around the origin is shown and all repeated segments are avoided for better visibility. The filled green and orange points are used to get the slope and the found tangent points are shown by open black circles.}
    \label{fgr:2D_f_effect}
\end{figure}

\begin{figure}
    \centering
    \includegraphics[width=0.8\textwidth]{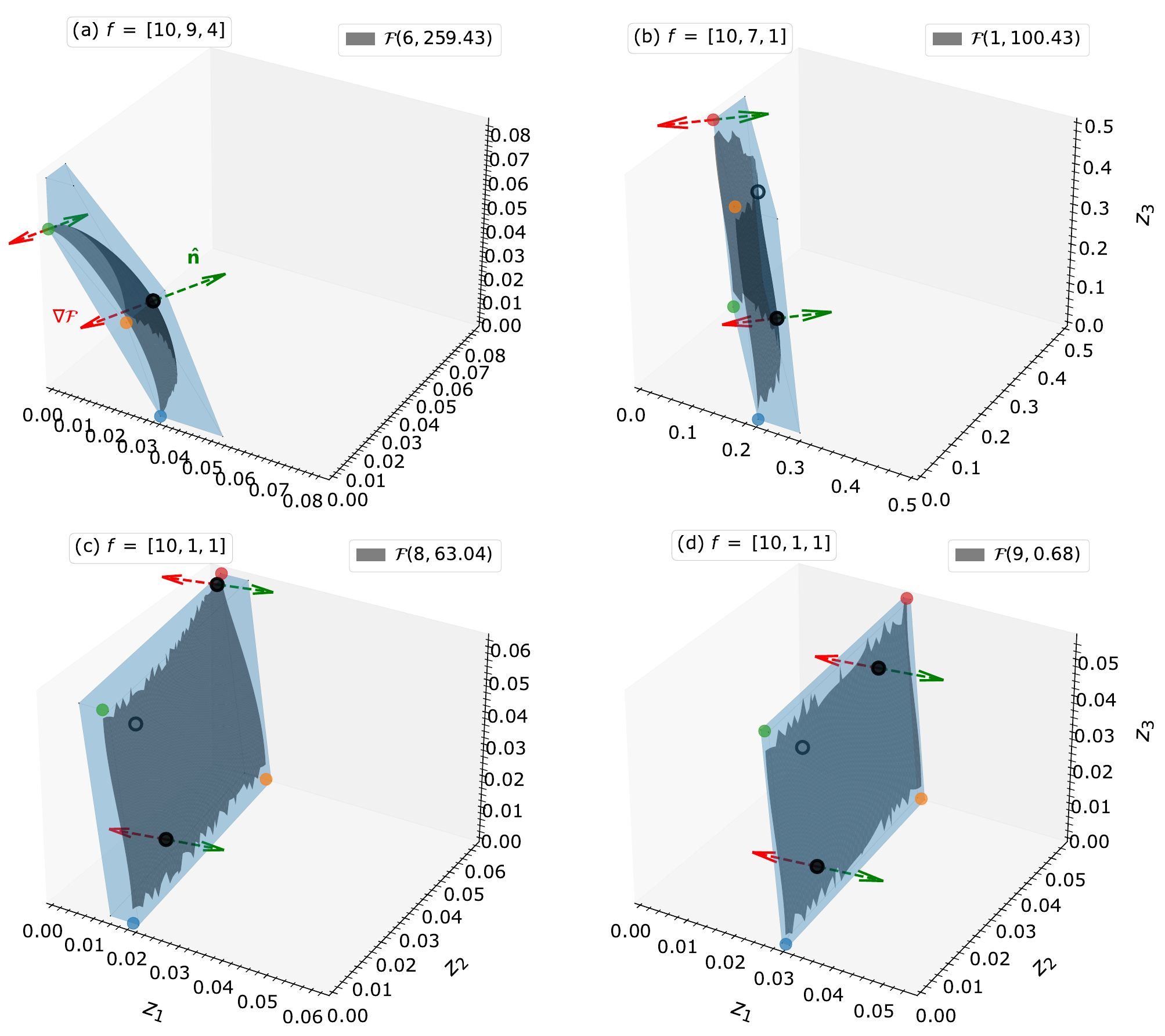}
    \caption{The different possible topologies of $\isof$ in $\P{3}$ on varying atomic scattering factors $f_i$ in Eq.~\eqref{eq:F}. The $\isof$ are differentiated according to their intersection along the $z_i$ directions. They can intersect (a)~with all three axes (closed topology), (b)~only with the two axes $z_1$ and $z_2$, or (c)~and (d)~only with axis $z_1$ (open topologies). The very first segment of the linearization around the origin is shown and all repeated segments are avoided for better visibility. Also, the computed gradient of $\isof$ and the normal vector are depicted at different tangent points. The filled, colored dots are used to define the required normal vector. The tangent points representing planes nearest to and farthest from the origin are represented by solid black dots, while the open black circle indicates an additionally computed tangent point.}
    \label{fgr:3D_f_effect}
\end{figure}

In general, the values of the atomic scattering factors are found to influence the curvatures of the isosurfaces. We learned that an anisotropic $\isof$ may show cases with an open topology along the direction where the less contributing $f_i$, \ie the smaller $f_i$, are assigned. For cases of similar atoms, where all structure factors $f_i$ are alike (\eg EPA model), closed topologies appear. Some of these isosurfaces have been exemplarily examined and are presented in~Fig.~\ref{fgr:2D_f_effect} and~\ref{fgr:3D_f_effect}. The different shapes result from varying ratios of the atomic scattering factors. 

A complete overview of the different isosurfaces~$\isof$ in $\P{2}$ is realized via keeping~$f_1$ at a constant value of $10$ and varying~$f_2$. As $f_2$ increases, the curvature of the isosurface changes, resulting in two categories of $\isof$: those that cut both main axes (closed topology) and those intersecting with only one main axis (open topology). Figure~\ref{fgr:2D_f_effect} shows that $\isof$ bends and tends to intersect both axes $z_1$ and $z_2$ as $f_2$ increases and becomes more similar to~$f_1$.

A similar study is carried out in~$\P{3}$ by assuming different $f_i$~combinations. Figure~\ref{fgr:3D_f_effect} summarizes the observed isosurfaces upon varying the ratios of the scattering factors $f_i$. As in $\P{2}$, the isosurface in $\P{3}$ may exhibit an open topology along the $z_i$ direction, which is associated with a low scattering factor $f_i$, \cf the open topology along $z_3$ in Fig.~\ref{fgr:3D_f_effect}(b), along $z_2$ and $z_3$ in Fig.~\ref{fgr:3D_f_effect}(d). The isosurface intersects all axes when all $f_i$ are considerably similar. 

In order to control the direction where the open topology may appear and to simplify the cases that may occur during the PSC approach, we sort the atoms according to their scattering strength in the initialization step such that the atom with the smallest scattering factor is associated with the highest index of~$z$.

\subsubsection{Generating the segments}\label{linearize-step1}

The linearization of any complete isosurface $\isog$ or $\isof$ starts with defining a segment in the vicinity of the origin. The concept of linearization can be easily understood in $\P{2}$ and can be systematically extended for higher dimensions. The isosurfaces for a given $l$ have a period of $1/l$, \ie each isosurface is repeated in each complete PS in translations of  $1/l$ along each axis. Hence $1$, $1/2$, and $1/3$ are the periods for $l~= 1$, $2$, and $3$, respectively, \cf Fig.~\ref{fgr:isoex1}(a)--(c). This characteristic period assists in the linearization. 

The first task is to find the coordinates $\bm{z}$, where the isosurface intersects with the axes in order to define the inner and outer boundaries. In $\P{2}$, this coordinate can be determined by rearranging Eq.~\eqref{eq:F} to~$z_2$, while $z_1$ is set to zero:

\begin{align}
    z_2 &= k^{-1} \cdot \arccos (\nicefrac{ s(l)\cdot|F_0(l)|}{2f_2} - \nicefrac{f_1\cos(k \cdot 0)}{f_2}),
\end{align}

where $k = 2\uppi l$, hereafter. In the case of $\P{3}$, setting two of $z_i$ (for example $z_1$ and $z_2$) to zero yields the remaining $z_j$ ($i \neq j$), \eg 

\begin{align}
    z_3 &= k^{-1} \cdot \arccos (\nicefrac{s(l)\cdot|F_0(l)|}{2f_3} - \nicefrac{f_1\cos(k \cdot 0)}{f_3} - \nicefrac{f_2\cos(k \cdot 0)}{f_3}).\label{eq:eq7}
\end{align}

These intersections with the main axes are represented by the filled points in Fig.~\ref{fgr:linearization}(b) and~(c). The determination of the boundaries for closed and open topologies (see Sec.~\ref{Topologies of isosurface}) differs significantly from each other. Therefore, we will discuss them separately.

\paragraph*{Closed topologies}

For closed topologies, all intersections with the axes exist and the inner boundary is devised by forming a line (in $\P{2}$) or a plane (in $\P{3}$) from those determined intersection points. The filled points in Fig.~\ref{fgr:linearization} define the inner boundary described by a unique normal vector $\bm{\hat{n}}$, which assists in determining the outer boundary.

The outer boundary is defined by employing a parallel shift of the inner boundary. Therefore it is crucial to determine the corresponding tangent point (open blue circles in Fig.~\ref{fgr:linearization}) on the isosurface with respect to the normal of the inner boundary. The line or plane formed at this juncture delineates the outer parallel boundary, marking the point beyond which $\isof$ ceases to exist. The existence of this specific tangent point is assured by the \textit{Mean Value Theorem}~\cite{Hobson1909}, which -- for the general non-EPA case -- states that a normalized gradient of the isosurface equivalent to the unit normal exists. With knowledge of $\bm{\hat{n}}$ from the inner boundary, a tangent point can be found by solving $\nicefrac{\nabla\mathcal{F}}{|\nabla\mathcal{F}|} =- \hat{\bm{n}}$; for derivations see Sec.~\ref{sec:scalarequ}.

\paragraph*{Open topologies}

For a given set of atoms with varying scattering contributions $f_i$ and a specific intensity or amplitude, the isosurfaces may show an open topology and not intersect with all the axes, see Sec.~\ref{Topologies of isosurface}. In such cases, the curvature of the isosurface is not convex over the full period but switches from convex to concave in the local vicinity of the open topology. In consequence, lines and planes derived from the intersections (closed circles in Fig.~\ref{fgr:linearization}) will cut $\isof$. Thus, for these cases, the determined lines/planes must be shifted both towards the origin and away from it to define limiting but not restricting boundaries. Appropriate anchor points for these shifts are defined with a similar strategy as for the closed topologies: We search for the tangent points at the isosurface with respect to the normal vector of the line/plane defined by the intersection points. Due to the change between concave and convex behavior, at least two such tangent points exist, which serve as anchor points for inner and outer boundaries, respectively.

\paragraph*{Finalizing the segment creation}

Once the inner and outer boundaries are defined, the segment can be created using the boundary distances between the origin and the inner and outer boundaries, respectively. For a given $l$, the boundary distances and $\bm{\hat{n}}$ are unique and together they provide \textit{segment descriptors}. This segment linearizes only a part of $\isof$ in the vicinity of the origin, \cf Sec.~\ref{Topologies of isosurface}. The remaining parts of $\isof$ are linearized by utilizing the translational and rotational symmetries as described in Sec.~\ref{Completing the linearization}. The routines are generally applicable, including the EPA case with geometric structure factor $G(l)$ and the corresponding isosurfaces $\isog(l,g_0(l))$. 

\subsubsection{Finding tangent points}\label{sec:scalarequ}

In the linear approximation, finding the tangent point is vital as well as critical. To find the tangent point, the least-squares (LS) refinement~\cite{dekking2005,lawson1974} is employed for the isosurface~$\isof$~Eq.~\eqref{eq:G}. Within the refinement, the deviation between a desired normal unit vector~$\hat{\bm{n}}$ and the direction of the gradient of~$\isof$ is minimized to identify the parallel hyperplane tangent to~$\isof$. Exemplarily for~$\P{2}$, a selected initial point is shifted along~$\isof$ to meet the requirement 

\begin{align}
    - \hat{\bm{n}} &= \frac{\nabla\mathcal{F}}{|\nabla\mathcal{F}|} \equiv \nabla \hat{\mathcal{F}}, \label{eq:Lag}
\end{align}

where the~$ \hat{\bm{n}} = \left(a_1,~a_2\right)$ has the fixed components $a_1$ and $a_2$ along the $z_1$ and $z_2$~axes, respectively. At the required tangent point, the conditions 

\begin{align}
     \nabla \hat {\mathcal{F}} \cdot \hat{\bm{n}} = \pm 1 \quad \text{and} \quad \left| F(l) \right|^2 - I = 0 \label{eq:corrdinate}
\end{align}

must be satisfied, with the sign $+1$ and $-1$ of $\nabla \hat{\mathcal{F}} \cdot \hat{\bm{n}}$ denoting the convex and concave curvature of $\isof$. These conditions define the error function to be solved numerically. The points $\bm{z}$, where $\isof$ cuts the axes (calculated from Eq.~\eqref{eq:eq7}, \cf the colored points in Fig.~\ref{fgr:3D_f_effect}), and the point $\bm{z}$ along the main diagonal (\ie the $z_i$'s of $\bm{z}$ are equal, determined from $z_i = k^{-1} \cdot \arccos \left(\nicefrac{|F_0(l)|}{ 2 \cdot \sum_{j=1}^{m} f_j }\right)$ with $s(l) = +1$) are considered as the different initial guesses for the possible tangent points to start the iterative process. Depending on the curvature of $\isof$, each initial guess can yield similar or different tangent points, which are dealt with differently for open and closed topologies.

The isosurfaces with closed topologies have a single convex curvature and will have a single tangent point, \cf Fig.~\ref{fgr:3D_f_effect}(a). Isosurfaces with open topologies may also show additional tangent points due to concave-convex curvature change, \cf Fig.~\ref{fgr:3D_f_effect}(b)-(d). These tangent points as well as the intersection points of~$\isof$ with the axes are used to determine the distance of the outer and inner boundaries. For the assumed case of $f = [10, 7, 1]$, the single open topology results in two different tangent points, \cf Fig.~\ref{fgr:3D_f_effect}(b). The tangent point closer to the origin gives the inner boundary distance, the point further away is used to fix the outer boundary. The isosurfaces with double open-topology result in more than two tangent points due to further changes in the curvature near all intersecting points, \cf Fig.~\ref{fgr:3D_f_effect}(c) and~(d).

The above formulations are followed for PS of dimension 3 or higher. However, the parallel boundary lines in $\P{2}$ offer an alternative elegant solution to solve for the tangent point. The normal vector $\bm{\hat{n}}$ in $\P{2}$ can be replaced by the slope $\zeta$ of the inner (and outer) boundary, which is defined as

\begin{align}
  \zeta &= \frac{\mathrm{d}z_2}{\mathrm{d}z_1} = \frac{ \mathrm{d}\left( k^{-1} \cdot \arccos (\nicefrac{|F_0(l)|}{2 \cdot f_2} - \nicefrac{f_1\cos(kz_1)}{f_2})  \right) }{\mathrm{d}z_1}.
\end{align}

Using the identity $\nicefrac{\mathrm{d}(\arccos(x))}{\mathrm{d}x} = \nicefrac{-1}{\sqrt{(1-x^2)}}$ and for simplicity applying $s(l) = +1$, the above equation becomes

\begin{align}
    \zeta &=\frac{-k^{-1} \left( k\cdot \frac{f_1\sin(kz_1)}{f_2}\right)}{ \sqrt{ 1 - \left( \frac{|F_0(l)|}{2 \cdot f_2} - \frac{f_1\cos(kz_1)}{f_2}\right)^2} }.
\end{align}

After resolving the root and rearranging the above equation, we receive

\begin{gather}
    \begin{aligned}
    (1-\zeta^2)\cdot \left(\frac{f_1}{f_2} \cdot \cos(kz_1) \right)^2+&\\
    2\zeta^2 \left( \frac{|F_0(l)|}{2 \cdot f_2} \right) \cdot \left(\frac{f_1}{f_2} \cdot \cos(kz_1) \right)+&\\
    \zeta^2 \left(1- \left( \frac{|F_0(l)|}{2 \cdot f_2} \right)^2 \right) - \left( \frac{f_1}{f_2} \right) ^2& = 0.
    \end{aligned}
\end{gather}

With $\xi = \frac{f_1}{f_2} \cdot \cos(kz_1)$ above equation becomes,

\begin{align}
    \resizebox{.9 \textwidth}{!}{%
    $(1-\zeta^2)\cdot \xi^2 + 2\zeta^2 \left( \frac{|F_0(l)|}{2 \cdot f_2} \right) \cdot \xi + \zeta^2 \left(1- \left( \frac{|F_0(l)|}{2 \cdot f_2} \right)^2\right) - \left( \frac{f_1}{f_2} \right) ^2 = 0.$%
    }
    \label{eq:18}
\end{align}

Solving this equation for $\xi$ and respectively $z_1$ based on the identified slope~$\zeta$ from the inner boundary gives the linear function for the outer boundary as well. In particular, the two roots of Eq.~\eqref{eq:18} can probe all possible maxima or minima of $\isof$. If both roots are valid, \ie real and smaller than~$1$, then $\isof$ will have two tangent points, \cf Fig.~\ref{fgr:2D_f_effect}(a)--(b), defining at the same time inner and outer boundaries. If only one root is valid or both roots are identical, then $\isof$ will have merely one tangent point defining the outer boundary, \cf Fig.~\ref{fgr:2D_f_effect}(c).

\subsubsection{Precision of the tangent point}\label{Exactness of tangent point}

Irrespective of topologies, the exact linearization of the isosurfaces is important to reduce false structure predictions or the number of pseudo-solutions. The quality of the linearization of $\isof$, particularly in higher dimensions $m \geq 3$, can be inferred from the metrics: (a)~the intensity at the tangent point and (b)~the angle between the normal $\bm{\hat{n}}$ and the gradient $\nabla\hat{\isof}$ of the isosurface. The found tangent point must be on the isosurface and hence must result in the same intensity value as that of $\isof$. The incorrect prediction of the tangent point may lead to differences between the structure-immanent solution space and the approximation, which may on the one hand unnecessarily increase the volume of the linearization segment and on the other hand, even more severe, excludes valid solution volume. The angle between $\bm{\hat{n}}$ and $\nabla\hat{\isof}$ should be \SI{180}{\degree}, due to the anti-parallel condition of Eq.~\eqref{eq:Lag} and we monitor the discrepancy by the \textit{deviation angle}. However, inherited from the mixed concave-convex curvatures of isosurfaces with an open topology, the intersecting points of $\isof$ on the main axes are no longer co-planar, inducing the need for approximation on defining $\bm{\hat{n}}$. We apply the \textit{Singular Value Decomposition} (SVD) method~\cite{Campbellactacrys} to determine the initial plane with the respective $\bm{\hat{n}}$ from the corner points (\eg the blue, green, red, and orange colored points in Fig.~\ref{fgr:3D_f_effect}(b)--(d)). The found~$\bm{\hat{n}}$ from SVD defines again the tangent point, the open black points in Fig.~\ref{fgr:3D_f_effect}.

In Tab.~\ref{tbl:Icompare}, both intensity and deviation angle are given to estimate the quality of linearization of respective isosurfaces for the challenging cases shown in Fig.~\ref{fgr:3D_f_effect}. For the given examples, the theoretical intensity (from the atomic structure) is identical to the one calculated from the tangent point, additionally the deviation angle is in the order of~$\num{e-4}$ in all cases.

\begin{table}
    \centering
    \caption{Comparison of theoretical (from the atomic structure) and calculated (from the tangent point) intensities for the atomic structure $[0.349, 0.362, 0.1615]$. The found normal vector $\bm{\hat{n}}$, the tangent point for the given atomic scattering factors $f_i$, and the deviation angles are also listed. A large value of the difference in angle indicates a larger deviation. The tangent point column lists two sets of coordinates in two rows corresponding to the inner (first row) and outer (second row) boundaries. The respective $\isof$'s along with the listed tangent points (filled black points) are given in Fig.~\ref{fgr:3D_f_effect}.}
    \resizebox{\textwidth}{!}{
    \begin{tabular}{c c c c c c c}
    \hline
    \hline
        Reflection & $f$ & \multicolumn{2}{c}{$I$} & $\bm{\hat{n}}$ & Tangent points & Deviation angle       \\ [1ex]
        &  & theoretical & calculated &  &  &  \\ [1ex]
		\hline     
        6 &  [10, 9, 4] & 259.427 & 259.427 & [0.680, 0.639, 0.358] & [0.000, 0.000, 0.063] &  0.0005  \\ [1ex]
          &             &         &         &                       & [0.018, 0.018, 0.027] &  0.0005  \\ [1ex]
        
        1 &  [10, 7, 1] & 100.431 & 100.431 & [0.777, 0.626, 0.055] & [0.000, 0.226, 0.500] &  0.0005  \\ [1ex]
          &             &         &         &                       & [0.144, 0.179, 0.093] &  0.0005  \\ [1ex]
        
        8 &  [10, 1, 1] & 63.042  & 63.042  & [0.992, 0.088, 0.088] & [0.002, 0.060, 0.060] &  0.0004  \\ [1ex]
          &             &         &         &                       & [0.017, 0.015, 0.015] &  0.0004  \\ [1ex]
          
        9 &  [10, 1, 1] & 0.68    & 0.68    & [0.996, 0.064, 0.064] & [0.023, 0.044, 0.044] &  0.0004  \\ [1ex]
          &             &         &         &                       & [0.029, 0.012, 0.012] &  0.0004  \\ [1ex]
        \hline
    \hline
    \end{tabular}
    }
\label{tbl:Icompare}
\end{table}

The current implementation still offers significant potential for enhancing the linearization method, specifically in resolving both inner and outer tangent points for cases involving open topologies and mixed concave-convex curvatures in higher-dimensional PS. One such enhancement could involve the development of a double-segment approach tailored for higher dimensions. Additionally, it is crucial that the determined boundaries fully encapsulate the entirety of the isosurface. To mitigate the risk of overlooking any valid solution spaces, we have introduced a cross-checking routine based on the grid-based method that verifies the complete enclosure of $\isof$ by the polytope. In the fail-safe case, the portion of the uncovered isosurface is returned to the momentary solution space, and the outer boundary distance is recalibrated, allowing for the continuation of the linearization process. However, it is important to note that this grid-based approach may demand substantial computational time, especially in higher-dimensional PS. Therefore, the development of a more efficient algorithm that guarantees the complete enclosure of $\isof$ while also validating the accuracy of the determined normal vectors remains a priority for future research and development efforts.

\subsubsection{Completing the linearization}\label{Completing the linearization}

The segment obtained in the previous step only linearizes a part of the isosurfaces with $s(l) = +1$. The complete $\P{m}$ encompasses $2\cdot l^m$ copies of this unique segment, due $l$-fold mirror symmetry along each $z_i$~direction: $l$~copies in all $m$~directions, maintaining a modulo of~$1/l$, see Fig.~\ref{fgr:isoex1}(a)--(c). Before performing this translational repetition in PS, the formed first segment is rotated around and mirrored about the origin to enclose further parts of the isosurface $\isog$ or $\isof$. 

The replication of the segment in our code is carried out using the mirror planes that pass through the origin. In $\P{2}$ there are two different mirror planes represented using the vectors $[1, 0]$, and $[0, 1]$ for $\isog$ and $\isof$ and two additional mirrors perpendicular to the main and secondary diagonals exclusively for $\isog$. The mirror planes represented by the vectors $[1, 0, 0]$, $[0, 1, 0]$, and $[0, 0, 1]$ are used for $\isog$ and $\isof$, as well as additional mirrors perpendicular to the main and secondary diagonal exclusively for $\isog$. The segment descriptors are then multiplied by these vectors to form further rotated segments in both closed and open topologies, \cf Fig.~\ref{fgr:G(3,1.56)} as an example for four/eight segments in $\P{2}$. This initial set of equivalent segments is then repeated in PS with translation vectors $\bm{R}$ as follows:

\begin{align}
    \bm{R} &= (\Delta v_1, \ldots, \Delta v_m), \label{eq:R} 
\end{align}

where $\Delta v_i$ are the translations along $z_i$~direction in~$\P{m}$, varying between \num{0} and \num{0.5} in increments of $\nicefrac{1}{2l}$. Those translations are applied only if the \textit{centers of polarity} condition at maximum amplitude is fulfilled, demanding that all signs of the individual contributions are equal. The \textit{centers of polarity} condition is defined as

\begin{align}
\big( \cos(k \cdot \Delta v_1) , \ldots , \cos(k \cdot \Delta v_m) \big) = 
    \begin{cases} 
        (+1,+1, \cdots ,+1), & \text{valid}, \\ 
        (-1,-1,\cdots,-1), & \text{valid}, \\ 
        \text{any other result} & \text{invalid}. 
    \end{cases}
    \label{eq:polarity}
\end{align}

A $+$ ($-$) center is found if all elements in Eq.~\eqref{eq:polarity} are positive (negative). Any other combination is denoted as mixed translation centers and describes positions, where there is no presence of maximum or minimum amplitude, \cf Fig.~\ref{fgr:G(3,1.56)}. For example, the combination $(\Delta v_1, \Delta v_2) = (0, 0)$ in $\P{2}$ has $+$ polarity, since $[\cos(0), \cos(0)] = [+1, +1]$. The combination $(\nicefrac{1}{6}, \nicefrac{1}{6})$ has $-$ polarity due to $[\cos(2 \uppi \cdot 3 \cdot \nicefrac{1}{6}), \cos(2 \uppi \cdot 3 \cdot \nicefrac{1}{6})] = [-1,-1]$. However, combinations such as $(\nicefrac{1}{2}, 0)$ have mixed polarity, \ie $[\cos(2 \uppi \cdot 3 \cdot \nicefrac{1}{2}), \cos(0)] = [-1, 1]$ and are invalid translation centers in $\bm{R}$ construction. Once the reflection~$l$ is defined, we can derive the translation vectors~$\bm{R}$ to ensure proper selection of $\Delta v_m$ values. These $\Delta v_m$ values are then utilized to replicate the segments and enclose the $2 \cdot l^m$ copies of $\isog$ or $\isof$ unique segments within~$\P{m}$. This iterative procedure effectively completes the linearization process.

\begin{figure}
    \centering
    \includegraphics[width=1\textwidth]{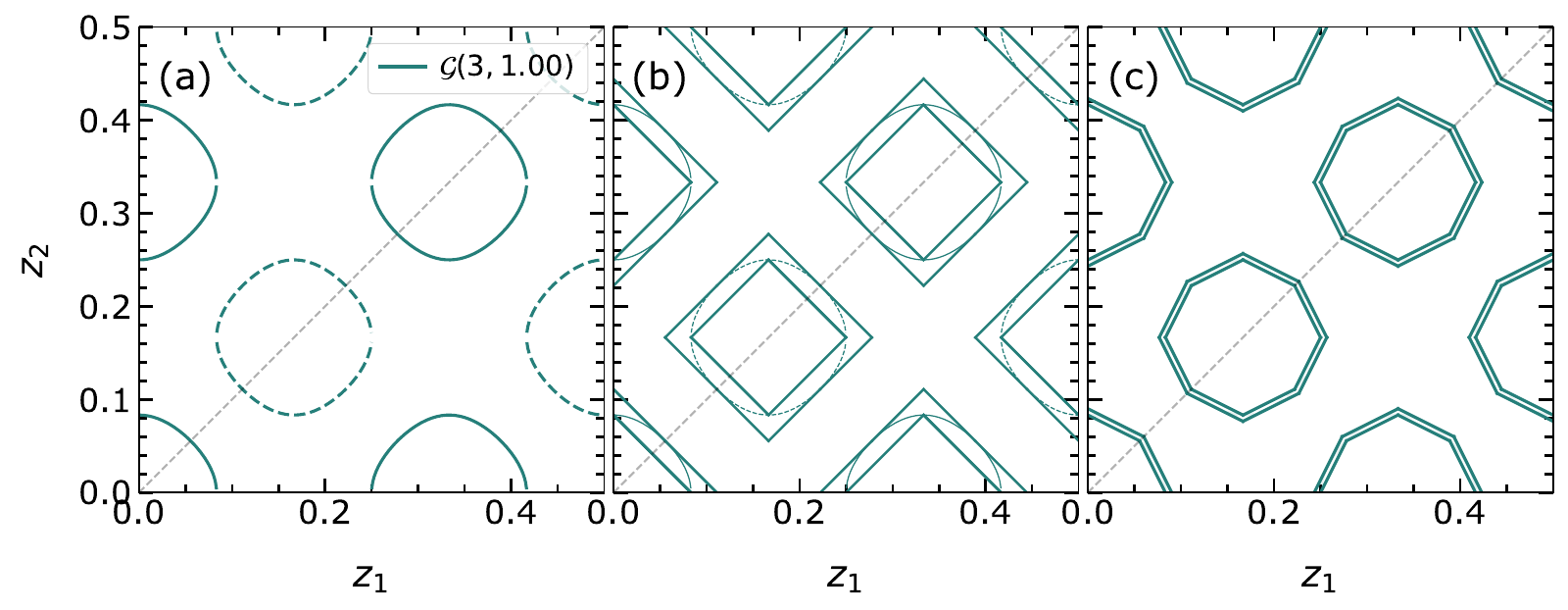}
    \caption{An example $\isog$ generated for $l = 3$ with $g(l) = 1.00$ in complete $\P{2}$. (a)~the $\isoG{3}{1.00}$ and their linearization (b)~with single-segment and (c)~with double-segment. The continuous and dotted lines in~(a) represent $s(l)=+1$ and $s(l)=-1$ respectively. They are centered around the points with '$+$' or '$-$' maximum amplitude.}
    \label{fgr:G(3,1.56)}
\end{figure}

\begin{figure}
    \centering
    \includegraphics[width=1\textwidth]{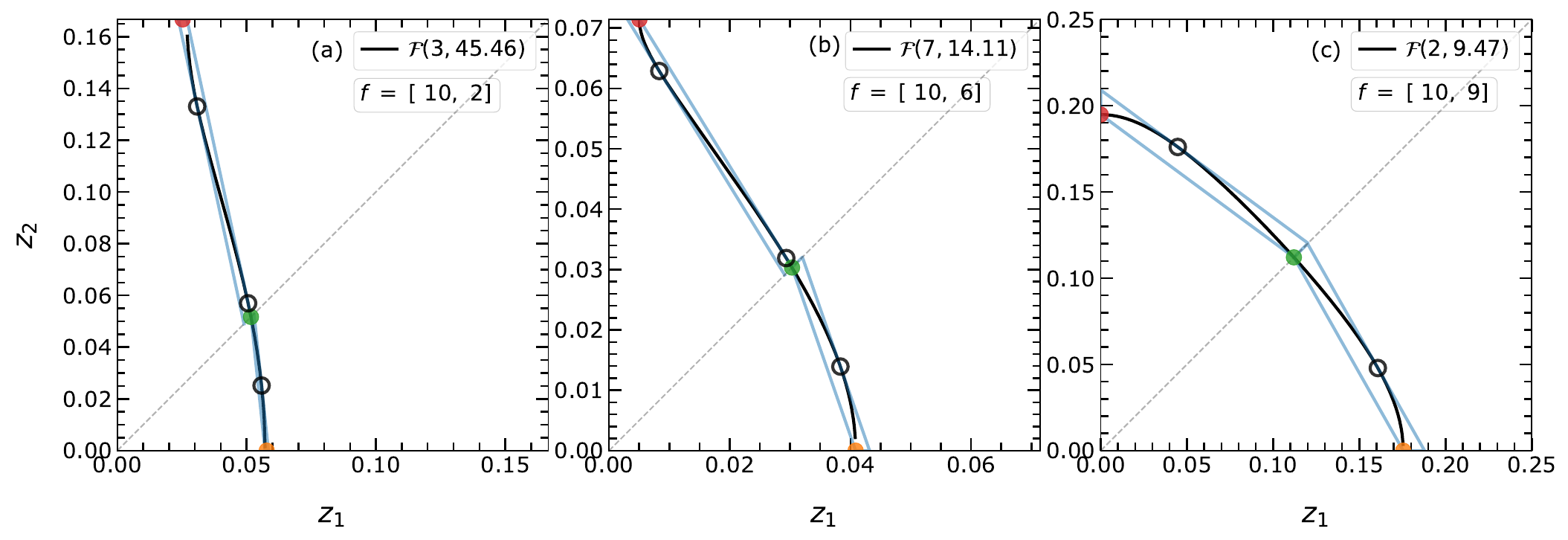}
    \caption{Application of the double-segment approach to the isosurfaces from Fig.~\ref{fgr:2D_f_effect}. The filled green-orange (green-red) points are used to calculate the slope for part of $\isof$ below (above) the diagonal line. The found tangent points are shown by open black circles. When the part of $\isof$ is a concave-convex mixed curvature, it possesses multiple tangent points for a given slope (\cf (a) and~(b)) and Eq.~\eqref{eq:18} is capable of exploring all points in a single analysis. In the case of multiple tangent points, all points are used to define inner and outer boundaries.}
    \label{fgr:2DDS_f_effect}
\end{figure}

\subsubsection{Improving the linearization}

The explained procedure of linearization for $\P{2}$ and $\P{3}$ involves the definition of a single segment that fully encompasses the isosurface in the vicinity of the origin. However, the defined segment includes a rather large solution space and, thus, exhibits a potential to minimize not only the unwanted solution space but also to propose a new linearization technique. This can be achieved by increasing the number of used segments. We will refer to this approach as the double-segment linear approximation. Figures~\ref{fgr:linearization}(a) and~(b) depict a comparison of these different linearization variations in $\P{2}$ for closed topologies. 

The transition from the single- to the double-segment approximation is achieved by dividing the PS into two parts in the most advantageous way: at the main diagonal, where the components of $\bm{z}$ are identical. The intersection of the isosurface with the main diagonal, which is represented by the filled orange point in Fig.~\ref{fgr:linearization}(a), \ie $z_1=z_2= k^{-1} \cdot \arccos (\nicefrac{ s(l) \cdot g(l)}{\sum f_i})$, serves as the first point to determine the inner boundaries of the two desired segments. Then setting $z_1$ or $z_2$ to zero results in the respective $z_2$ or $z_1$ component of the second point (filled green points in Fig.~\ref{fgr:linearization}(a)). The two straight lines formed from these two points with the orange point define the inner and outer boundaries in a similar way as for the single-segment approach. Following the same methodology, the double-segment approximation is successfully developed also for non-EPA, only in $\P{2}$ PS. As depicted in Fig.~\ref{fgr:2DDS_f_effect}, the isosurfaces from Fig.~\ref{fgr:2D_f_effect} are approximated using double-segments.

This improvement in linearization results in a significantly reduced solution space at the cost of additional computational load for handling a larger amount of polytopes, cf. Sec.~\ref{sec:2DMC-epa}~and~\ref{sec:2DMC-time}. However, it is important to note that the variation in the approximation is exclusively developed for~$\P{2}$ at the moment, while the single-segment linearization is employed for three- or higher-dimensional PS by default. Developing optimized variations other than single-segment linearization in higher dimensional PS is an important future goal and will be considered in future work.
 
\subsection{Intersection of linearized isosurfaces: The solution}\label{sec:solution}

Once the linearization is completed, the polytopes are created by collecting all segments for a given reflection to search for the solution region. The set of polytopes contains $2\cdot l^m$ segments for a given reflection~$l$. In principle, exactly one of the manifolds of an isosurface of reflection~$l$ always contains the crystal's structure solution, and therefore their corresponding segments also do. In turn, we need to find that region that is common to all the isosurfaces of different reflections; this process corresponds to determining the intersection between the polytopes.

The solution is found systematically through the following procedure. The solution space is reduced by intersecting the polytopes of the different reflections. Hence, with the addition of more reflections to the calculation, the solution space as well as the positional errors are in general gradually decreasing. This process is repeated until the last available reflection. Often, the final solution region consists of several very small disjoint areas that are often clustering. Subsequently, the final solution regions are used to calculate the coordinates of atoms in the structure of interest; these coordinates are presented as a list of atomic positions with their errors; as usual in crystallographic information file format. To determine the atomic position and the error in the analysis, we utilize the centroid of the solution regions and the extension in the $z_i$~directions. The extension of each component can be quantified as 

\begin{equation}
   \Delta z_i = z_i^{\max} - z_i^{\min}
   \label{eq:error}
\end{equation}

using the minimum $z_i^{\min}$ and maximum $z_i^{\max}$ values. These set operations are carried out by means of the \textit{polytope}~\cite{polytope} and \textit{shapely} packages~\cite{shapely}. The polytopes shown in all figures are generated with the~\textit{IntvalPy} package~\cite{intvalpy}.


\section{Results}\label{sec:results}

In the upcoming sections, we present an introductory example (Sec.~\ref{sec:example_2D}) as well as results of meticulous \textit{Monte-Carlo}~(MC) simulations of PSC to scrutinize the theory's performance through a large number of simulations in $\P{2}$ (Sec.~\ref{sec:2DMC}) and $\P{3}$ (Sec.~\ref{sec:3DMC}). The examples will cover EPA as well as non-EPA calculations as the variation of the structure factor can have a severe impact on the shape of the isosurface and thus on the linearization process and the validity of a determined solution concerning real data.

The Monte-Carlo simulations are analyzed to gain insights into the overall solution landscape. The volume of the solution regions and the error are monitored in each addition of reflections to pin any artifacts. For the given coordinate, we calculate and sum up the volumes of all obtained solution regions as simple performance descriptors. This cumulative volume is then used to construct a virtual, representative, $m$-dimensional sphere. Ideally, we receive a singular final solution identified within the overall solution landscape. However, in the case of numerous solutions, all are equally probable within the limits of linearized polytope regions, highlighting the significance of considering all possible solutions in the analysis. 

\subsection[Explanatory example in P2]{Explanatory example in $\P{2}$}\label{sec:example_2D}

\begin{figure}
    \centering
    \includegraphics[width=1\textwidth]{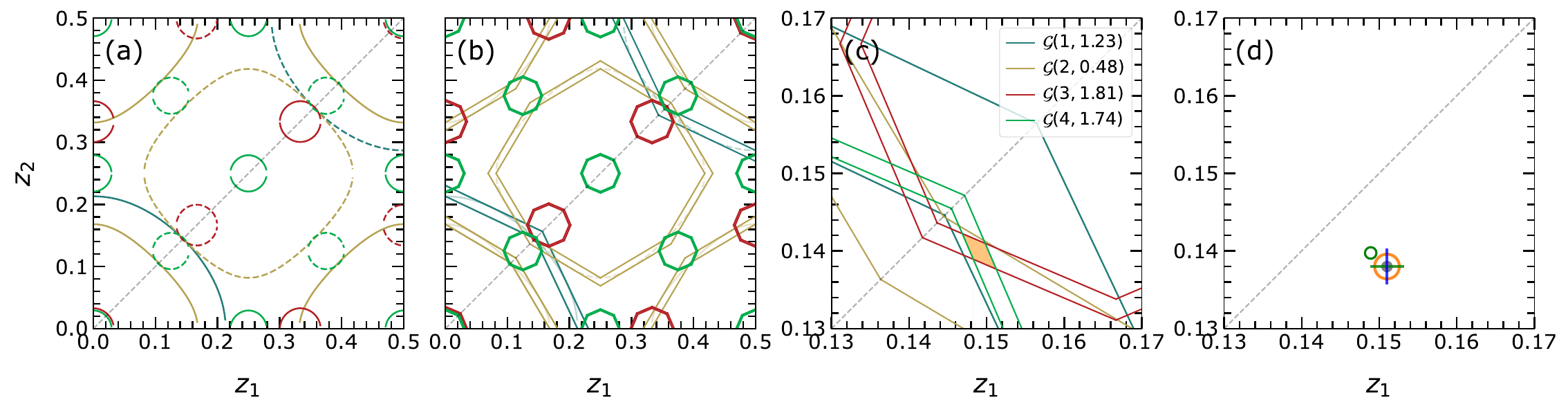}
    \caption{As a test case the coordinates $z_1 = 0.151$ and $z_2 = 0.138$ are solved by double-segment linear approximation and using the first four reflections for the EPA case and intensities with (a)~the $\isog$ for $l = 1,...,4$, (b)~the boundaries of linearization repeated in complete $\P{2}$, and (c)~the common intersection region from the linearized polytope with a total area of \num{7.99e-6}. The dashed gray lines denote the mirror symmetry in $\P{2}$ for EPA cases. The green open circle in (d)~denotes the centroid of the solution region (the shaded area shown in (c)) and is predicted to be $(z_1, z_2) = (0.1489 \pm 0.0021, 0.1397 \pm 0.0022)$. The green and blue bars represent the calculated errors in $z_1$ and $z_2$ and are attached to the test structure's assumed coordinates for reasons of comparability. A radius corresponding to the polytope area is calculated and represented as an open orange circle in (d).}
    \label{fgr:ex1}
\end{figure}

The effectiveness of the linear approximation is evaluated by solving an example structure consisting of two atoms, with coordinates $z_1 = 0.151$ and $z_2 = 0.138$. The structure is solved by considering the reflections \numrange{1}{4} whose intensity is converted into its associated $\isog$ within the EPA framework, \ie the scattering factor of two atoms is set to~1. Figure~\ref{fgr:ex1}(a) and (b) illustrate the superimposition of all $\isog$ and their linear approximations respectively.

The obtained isosurfaces~$\isog$ exhibit a predominantly smooth curvature and have a nearly circular topology in~$\P{2}$ (Fig.~\ref{fgr:ex1}(a)). The behavior is trivial due to the application of the EPA model for intensities. The determined solution in $\A{2}$ is $(0.1489 \pm 0.0021, 0.1397 \pm 0.0022)$, shown in Fig.~\ref{fgr:ex1}(c). The error is in the order of $\approx\num{e-3}$, which can be reduced further by increasing the number of reflections, for example, \cf Sec.~\ref{sec:2DMC-epa}. The solution is unique in $\A{2}$ (as shown in Fig.~\ref{fgr:ex1}(c)), however the mirror symmetries along the main diagonal lead to three additional, but equivalent solution regions in $\P{2}$.

If experimental errors are considered, they enlarge the width of the linearization polytopes and further distinct solution regions may appear~\cite{Zschornak2023}. The results are summarized in Fig.~\ref{fgr:ex1}(d) with the centroid (green circle), area (orange circle), and extent of the solution region (crossbar). Including more reflections in the calculation may merge the green and orange circles, reducing error and further decreasing the area of the solution region. The schematic as in Fig.~\ref{fgr:ex1}(d) is used to outline the results from the MC simulation in the sections below. 

\subsection[Monte-Carlo simulations in P2]{Monte-Carlo simulations in $\P{2}$}\label{sec:2DMC}

The performance of the developed linearization technique and the accuracy of the PSC approach are analyzed through MC simulations. Within all PSC routine variations, the same set of random atomic coordinates has been used to ensure comparability. The trend of area or volume (in $\P{2}$ or $\P{3}$, respectively) and extension of the solution region is monitored to track the deviations in structure prediction and identify weaknesses of the routines. Finally, the computing time is measured and presented for each MC simulation. The results are presented in the same style as in Fig.~\ref{fgr:ex1}(d) with an additional open yellow circle highlighting the total number of solutions. This yellow circle is centered in the same way as the orange one and its thickness changes according to the amount of identified total solutions. Hence, the thicker the yellow circles, the greater the number of solutions obtained. Also, timings and code performance for the MC simulations are evaluated in detail in Sec.~\ref{sec:2DMC-time}.

\subsubsection{\label{sec:2DMC-epa}MC simulation within EPA framework}

At first, randomly generated coordinates are solved using the EPA model, \ie the atomic scattering factors $f_i = 1$. Since the sign and magnitude of the amplitude $\isog$ are calculated explicitly, the given atomic structures can be solved using the following two different approaches. For the amplitude approach~$(2 g(l), \left|F_0(l)\right|)$, both sign and magnitude are utilized, and only the polytopes corresponding to the correct sign are considered in the calculation. In contrast, for the intensity approach~$(\left|2 g_0(l)\right|^2, \left|F_0(l)\right|^2)$ all polytopes are considered without the knowledge of the sign. The amplitude approach is particularly beneficial when the experimental data contains the sign.
 
\begin{figure}
    \begin{minipage}{14cm}
        \begin{overpic}[width=1.1\textwidth]{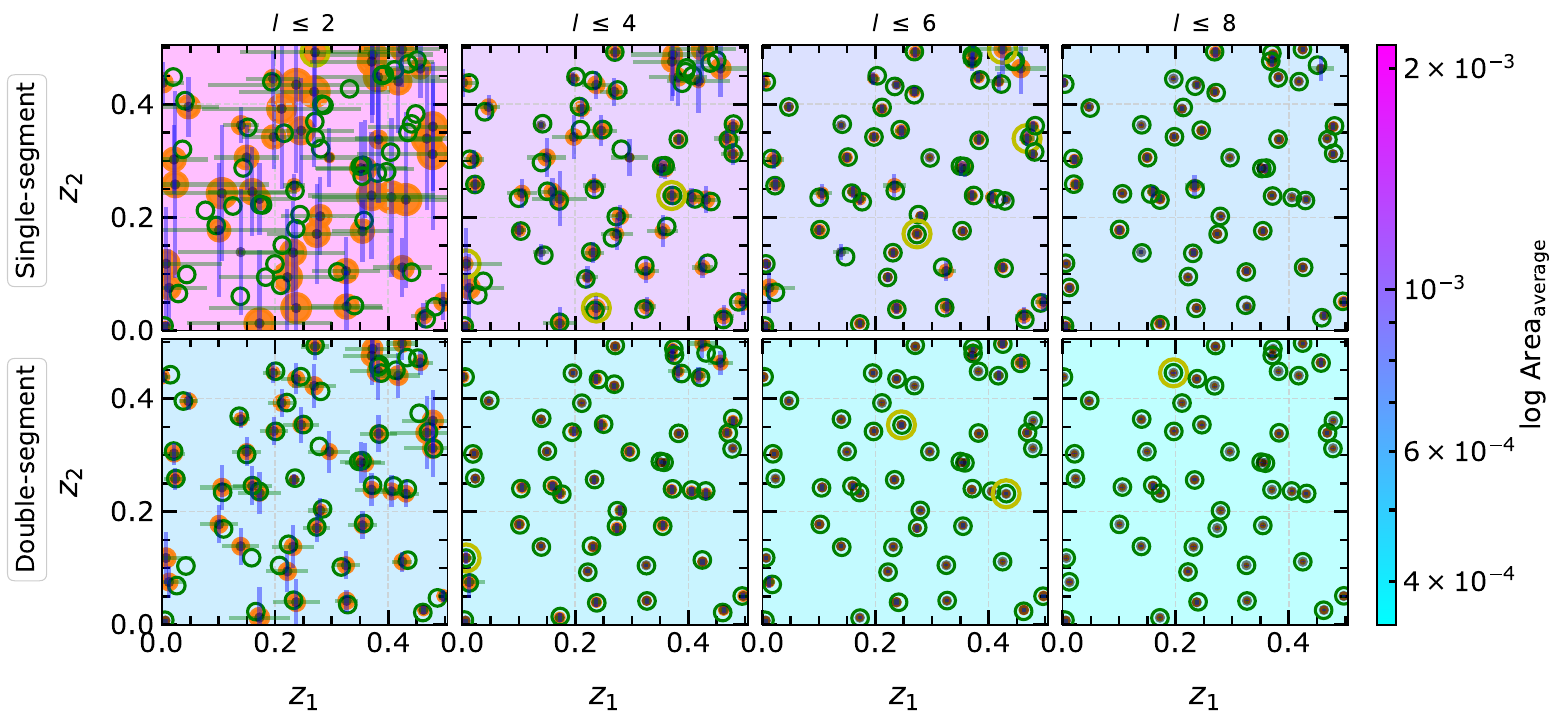}
            \put(00,7.4cm){(a) Amplitude approach}
        \end{overpic} 
    \end{minipage}\\\vspace*{1cm}%
    \begin{minipage}{14cm}
        \begin{overpic}[width=1.1\textwidth]{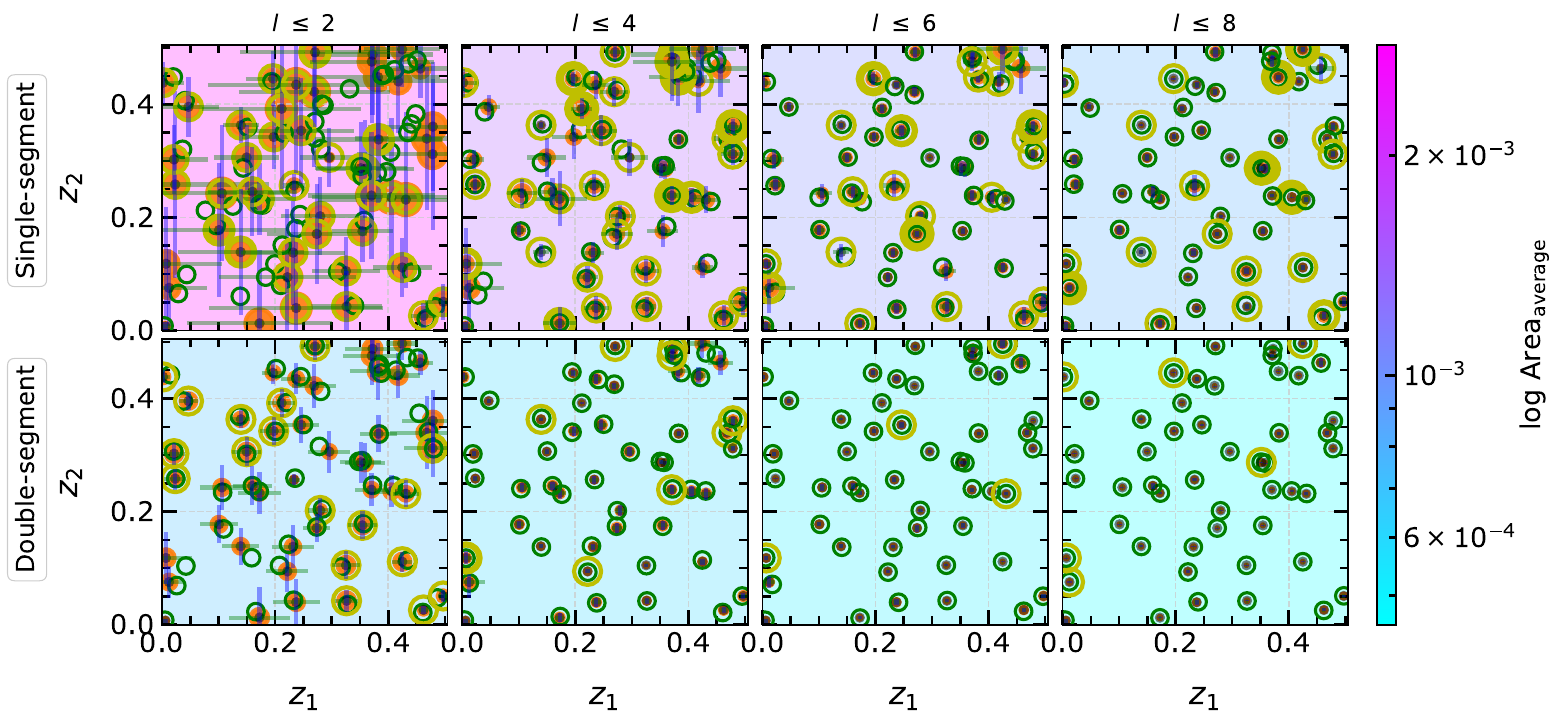}
            \put(00,7.4cm){(b) Intensity approach}
        \end{overpic}
    \end{minipage}
    \caption{Results of MC simulation using single- and double-segment linearization with amplitude and intensity approaches. The green circles surround the centroid of the solution region within the EPA framework. When more than one solution region is available, the number is represented by the yellow circle (increased thickness for a larger number, \cf Fig.~\ref{fgr:2D_Totalt}). The green and blue bars are the positional uncertainties on the computed $z_1$ and $z_2$, respectively. The area of all solution regions is summed up to calculate the radius of the virtual circle as $\sqrt{\nicefrac{\text{area}}{\uppi}}$, shown by the orange circle. The background color is defined by the average area of the solution regions of all random pairs, normalized to the overall minimum and maximum for all presented settings.}
    \label{fgr:2DMC_EPA}
\end{figure}

Figure~\ref{fgr:2DMC_EPA}(a) and~(b) illustrate the progression of the intersection region using both single- and double-segment approaches. As more reflections are added incrementally the positional uncertainties and the area of the solution region are continuously decreased. Figure~\ref{fgr:2DMC_EPA} shows that the double-segment linearization reduces the error more than the single-segment linearization. In comparison, the double-segment provides higher resolution by a factor of 2.7 with just the first two reflections and 3.6 with 8 reflections. The main reason is that the polytopes generated in single-segment include a larger area, thus the intersection regions tend to remain larger than those from the double-segment approach. Consequently, the positional uncertainties on the computed $z_1$ and $z_2$ are higher in single-segments. However, the area and positional uncertainties of the solution region exhibit steady improvement as more reflections are considered. Remarkably, with only four reflections, the double-segment linearization reduces the average area of the solution region below $10^{-3}$, while the single-segment linearization achieves a comparable result after considering as many as eight reflections. These findings highlight the advantages of employing multiple segments in enhancing the accuracy and precision of the computed atomic coordinates.

The amplitude and intensity approaches solve the coordinates, resulting in a similar real solution. In most cases, the amplitude approach results in a unique solution in $\P{2}$. However, the intensity approach induces many equivalent mirror solutions due to ambiguity in the sign, \cf the difference of thickness of yellow circles between Fig.~\ref{fgr:2DMC_EPA}(a) and~(b). Hence, as known from conventional XRD refinements, it is beneficial to measure both magnitude and sign for a unique structure prediction.

\subsubsection{MC simulation within Non-EPA framework}\label{sec:2DMC-nepa}

\begin{figure}
    \begin{minipage}{12cm}
        \begin{overpic}[width=\textwidth]{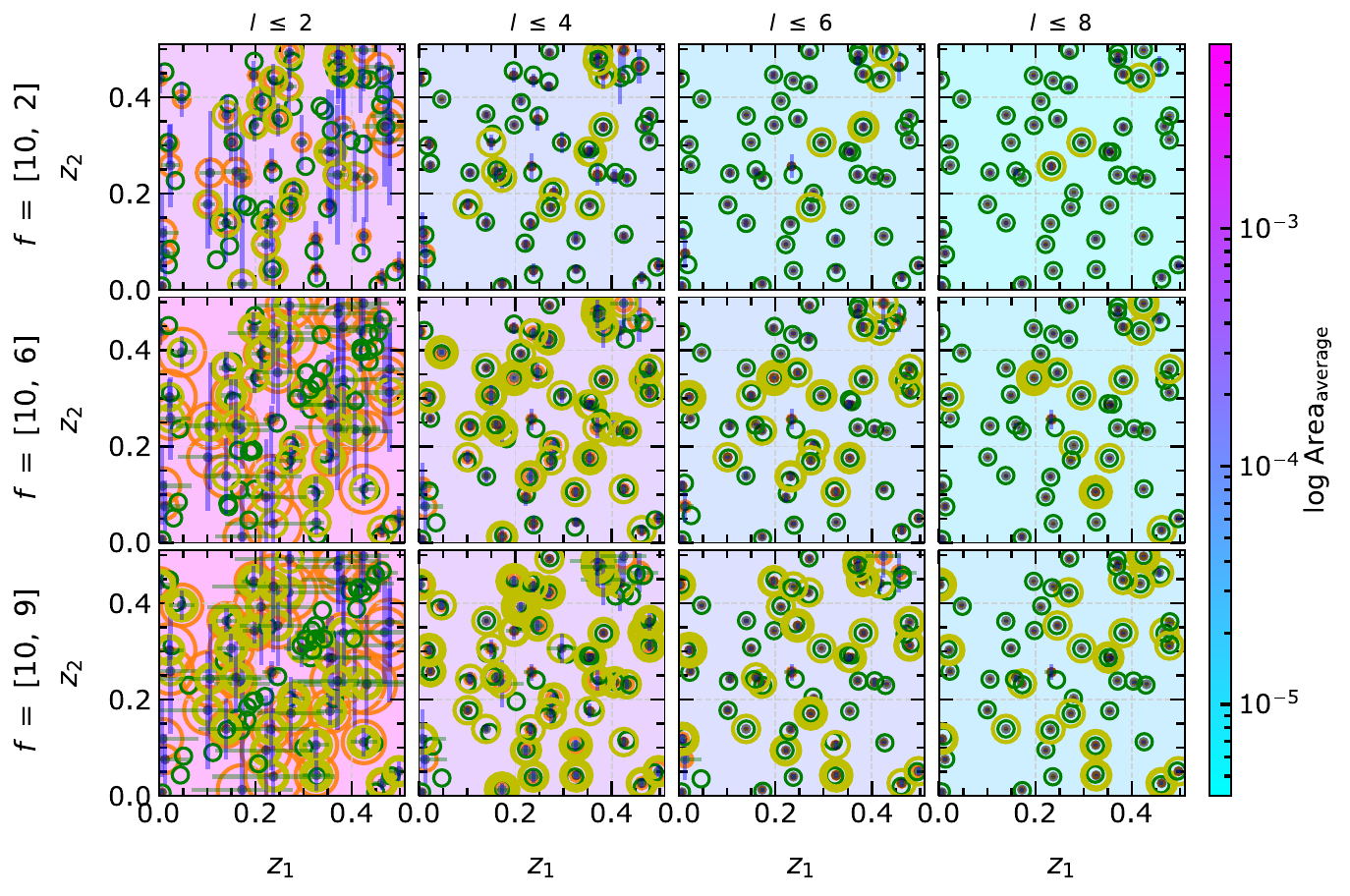}
            \put(00, 8.2cm){(a) Single-segment } 
        \end{overpic} 
    \end{minipage}\\\vspace*{0.5cm}%
    \begin{minipage}{12cm}
        \begin{overpic}[width=\textwidth]{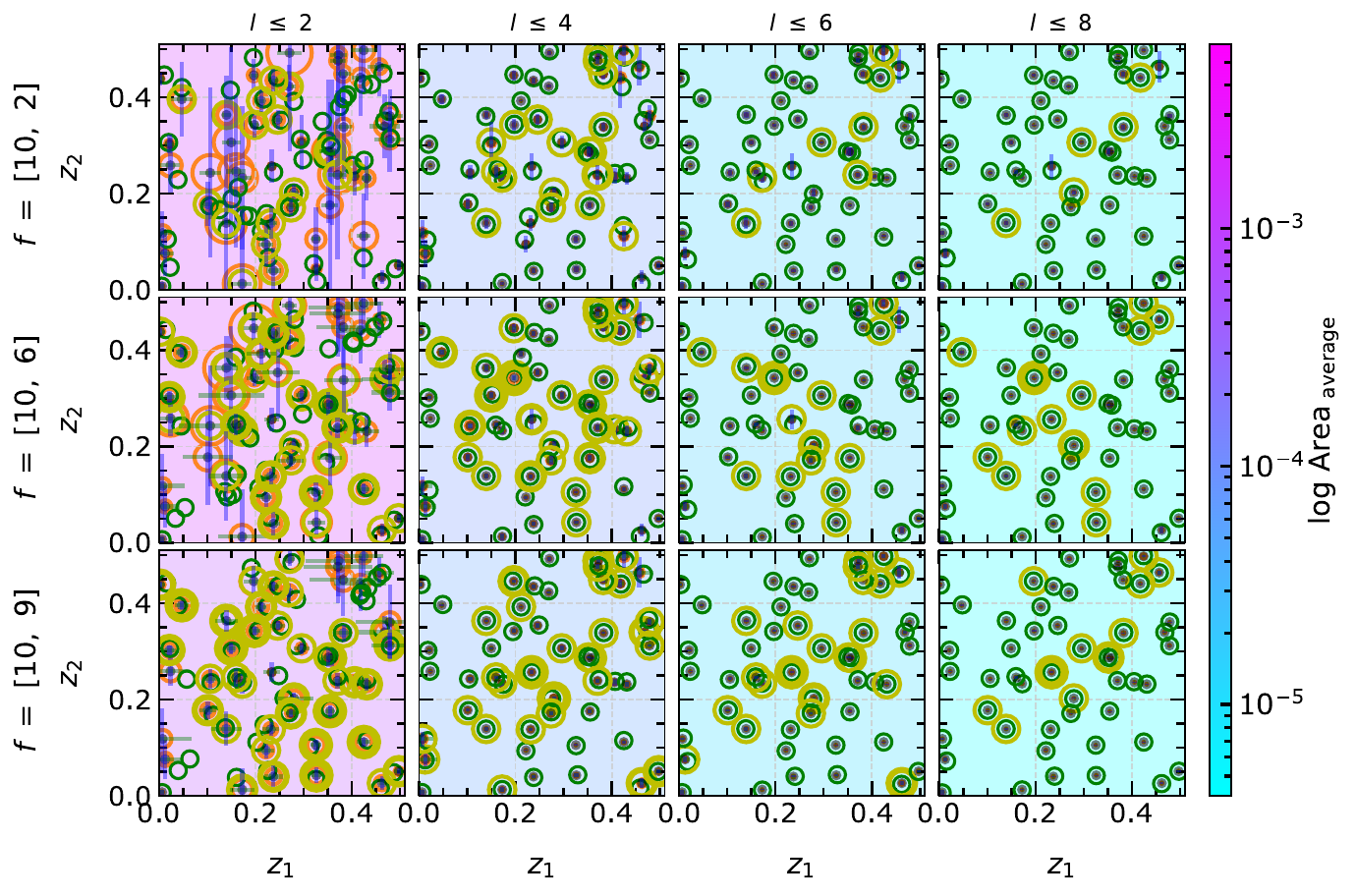}
            \put(00, 8.2cm){(b) Double-segment}
        \end{overpic}
    \end{minipage}%
    \caption{Solved atomic coordinates for single- and double-segment approaches using intensities within the non-EPA framework. The atomic scattering factors $[f_1, f_2]$ in Eq.~\eqref{eq:F} are set to $[10, 2]$ (top row), $[10, 6]$ (middle row), and $[10, 9]$ (bottom row) respectively to represent heavy-light, heavy-medium, and similar atom combinations along $z_1$ and $z_2$~directions. The results are shown for the increasing number of reflections involved in the solution-finding process. The black dots, green circles, and yellow circles (increased thickness for a larger number, \cf Fig.~\ref{fgr:2D_TotaltnEPA}) represent the generated random atomic coordinates, the centroid of the solution region, and the number of solution regions obtained respectively.}
    \label{fgr:2DMC_nEPA}
\end{figure}

The interesting question for realistic structures and diffraction data is to investigate the applicability of PSC for realistic atomic scattering factors $f_i$, which influence the solution-finding process significantly. The results including the heavy -- light atom combination $f = [10,2]$ as well as the combination of similarly weighted atoms $f = [10,9]$ are presented in Fig.~\ref{fgr:2DMC_nEPA}. For all the non-EPA cases, we focus exclusively on the more complex intensity approach, since it is the general case of experimental diffraction data with a higher multiplicity of solution regions and thus more demanding for the code.

The area of the solution region is set as the common scale across all subfigures in order to compare the effect of increasing the number of reflections in the calculation. Figure~\ref{fgr:2DMC_nEPA} demonstrates the crucial role of the $f_i$ ratios in defining the size of the solution region. The different $f_i$ cause anisotropically deformed $\isof$, \textit{\cf} Fig.~\ref{fgr:2D_f_effect}. In some cases, the linearization of such deformed $\isof$ for given $l$ may contain a significantly larger area than that from other $l$. This is directly reflected in the count of observed solution regions, denoted by yellow circles in Fig.~\ref{fgr:2DMC_nEPA}. Irrespective of these effects, the area of the solution region is again continuously reduced by increasing the reflection index in the calculations. 

The solutions of the calculations with $l \leq 2$ exhibit large positional uncertainties and areas, as indicated by the size of the orange circles in Fig.~\ref{fgr:2DMC_nEPA}. When considering up to 8~reflections, all atomic coordinates are successfully determined with positional uncertainties below $10^{-3}$ regardless of the specific combinations of~$f_i$. Additionally, Fig.~\ref{fgr:2DMC_nEPA} indicates that the PSC implementation effectively handles any possible combination of~$f_i$ in~$\P{2}$. Overall, the findings demonstrate the robustness and reliability of the PSC method in accurately determining atomic coordinates, even when dealing with diverse $f_i$ combinations.

\subsection[Monte-Carlo simulations in P3]{Monte-Carlo simulations in $\P{3}$} \label{sec:3DMC}

Following the MC simulations in $\P{2}$, the analogue investigations are carried out in the three-dimensional PS $\P{3}$. The polytopes are analyzed with their volume and extension in $z_1$, $z_2$, and $z_3$ directions. Again, the identified solutions by the code are counted and their volumes are summed up to construct a virtual sphere. So far, no double-segment linearization has been established for $\P{3}$, and thus we focus only on single-segment linearization. Timings and code performance for the MC simulations are evaluated in detail in Sec.~\ref{3DMC-time}.

\subsubsection{MC simulations within the EPA framework}\label{sec:3DMC-epa}

\begin{figure}
    \begin{minipage}{14cm}
        \begin{overpic}[width=\textwidth]{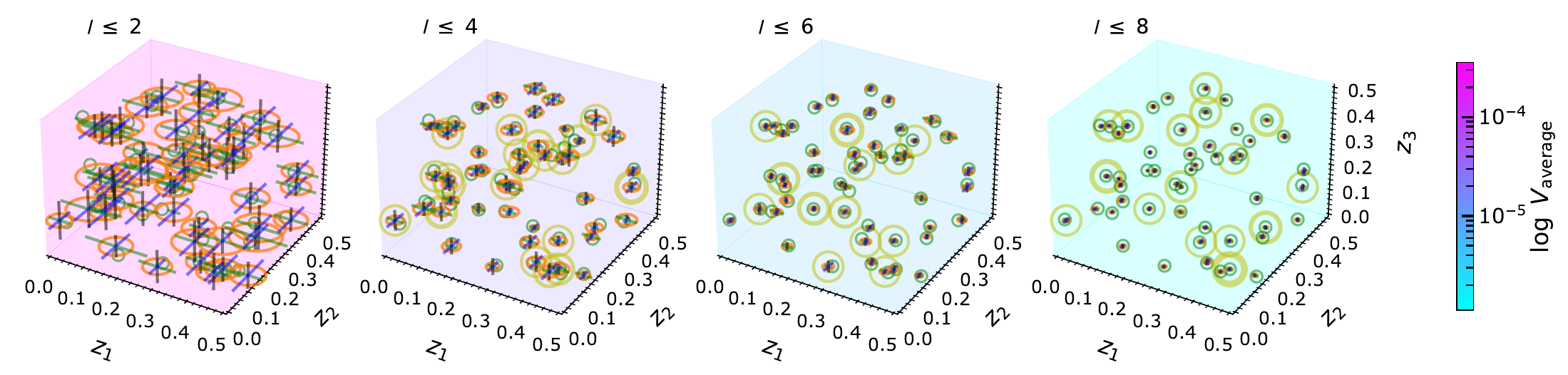}
            \put(00, 3.5cm){(a) Amplitude approach}
        \end{overpic} 
    \end{minipage}\\\vspace*{1 cm}%
    \begin{minipage}{14cm}
        \begin{overpic}[width=\textwidth]{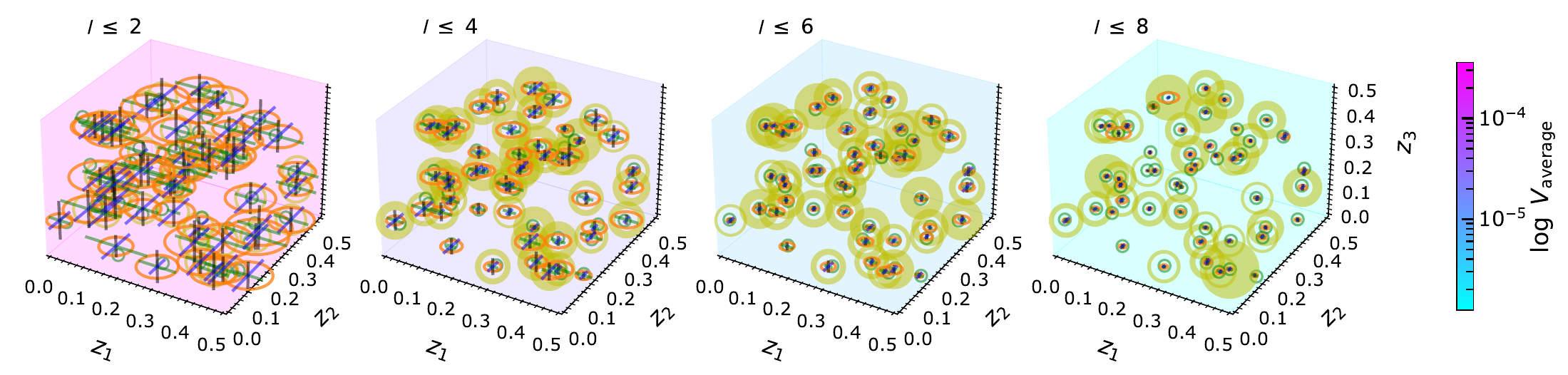}
            \put(00, 3.5cm){(b) Intensity approach}
        \end{overpic}
    \end{minipage}
    \caption{Results of MC simulations for $m=3$ within the EPA framework. The generated atomic coordinates are solved using amplitude (top row) and intensity (bottom row). The green circle represents the centroid of the polytope that encompasses the given atomic coordinates. The crossbar on each point indicates the error along each $z_i$ direction. When more than one solution region is available, the number is represented by the yellow circle (increased thickness for a larger number, \cf Figs.~\ref{fgr:3DMC_time_EPA} and \ref{fgr:3DMC_time_nEPA}). The volume of all found solutions $V$ is summed up to calculate the radius of the virtual sphere as $\mathcal{R}~=~\sqrt[3]{\nicefrac{3V}{4\uppi}}$ which is represented by orange circles. The background color represents the average polytope volume of all solution regions of all considered coordinates.}
    \label{fgr:3DMC_EPA}
\end{figure}

Figure~\ref{fgr:3DMC_EPA} presents the results of the Monte-Carlo simulations in~$\P{3}$ within the EPA framework, which uses both, amplitude and intensity approaches. The conclusions drawn from the $\P{2}$~case apply similarly to~$\P{3}$. Since the single-segment linearization includes a volume around the isosurface, the intersection process potentially reveals more than one solution. By gradually adding further reflections, the positional uncertainties and the size of the solution regions.

Our findings demonstrate that our code efficiently reproduces the given coordinates for the three structural degrees of freedom, regardless of the specific set of structural positions. Again, as observed in $\P{2}$, the intensity approach results in more possible solutions than the amplitude approach due to the ambiguity in the sign, and only a few solutions are identified uniquely. 

\subsubsection{MC simulation within non-EPA framework}\label{sec:3DMC-nepa}

\begin{figure}
    \centering
    \includegraphics[width=1\textwidth]{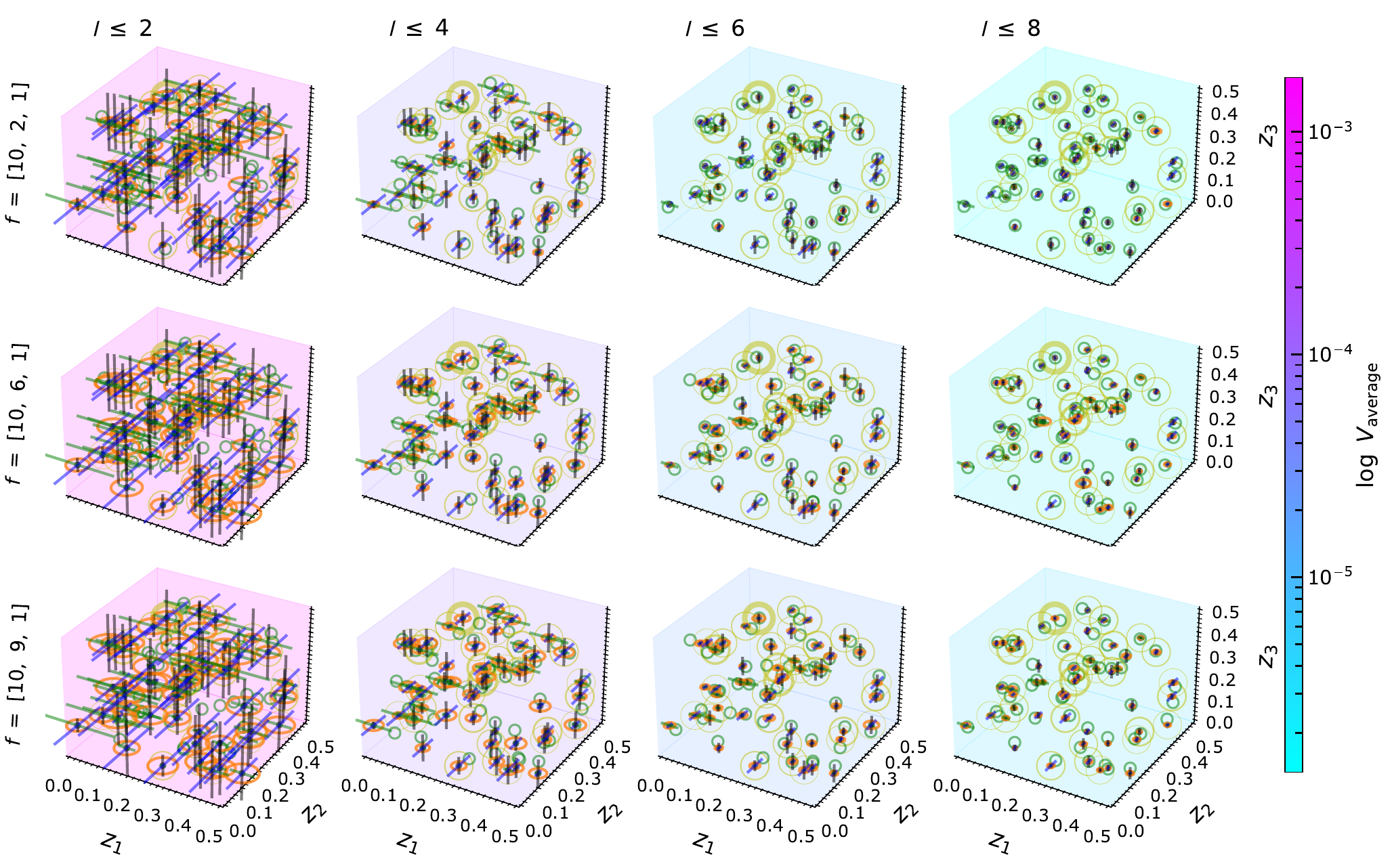}
    \caption{Results of MC simulations for $m=3$ within the non-EPA framework with intensity approach. The structures are solved for different settings of $f_i$. For the definition of different colors and symbols refer figure caption in Fig.~\ref{fgr:3DMC_EPA}.}
    \label{fgr:3DMC_nEPA}
\end{figure}

The MC simulations are repeated for general $f_i$ values (non-EPA) in $\P{3}$. As in $\P{2}$ non-EPA, different combinations for atomic scattering factors are considered in~$\P{3}$, covering the three scenarios of differently weighted atoms: heavy-light-light, heavy-medium-light, and heavy-heavy-light. We fixed $f_1$ and $f_3$ to $10$ and $1$, respectively, to represent heavy and light atoms, and only changed the contribution of $f_2$. These $f_i$ combinations are treated within the intensity approach. For comparability of the results, the initial atomic coordinates are kept identical for all scenarios, only the scattering factors~$f_i$ are varied, see Fig.~\ref{fgr:3DMC_nEPA}.

For each scenario of $f_i$~combinations, the expected reduction in volume and positional uncertainty of the solution region is visible, in analogy to the $\P{2}$ case. The variation in the $f_i$ ratios affects mainly the number of solutions as shown by the size of the virtual sphere in Fig.~\ref{fgr:3DMC_nEPA}. The curvature of $\isof$ varies strongly depending on $f_i$~values. These multiple solutions can be further minimized or eliminated by increasing the number of reflections in the calculation and by reducing the linearization volume. If the newly considered reflection does not result in a polytope with a smaller intersection volume than the previous reflection the intersection process will yield the same outcome as before. The presented non-EPA results demonstrate the capability of PSC in generally handling crystal systems with three structural degrees of freedom and different $f_i$ combinations, which can be utilized further for realistic diffraction data. 

\subsection{Timing benchmark}

As explained in Sec.~\ref{sec:steps}, the structure determination process consists of four distinct steps: initialization, linearization, intersecting, and writing. We monitor the time consumption of each step to benchmark the performance of our developed algorithm. The initialization step consumes a small amount of time, taking less than a millisecond. During the linearization step, we identify the first segment, which is then repeated in the complete $\P{2}$/$\P{3}$ with the constraint given by Eq.~\eqref{eq:polarity} as explained in Sec.~\ref{Completing the linearization}. This step consumes significantly more time. The details of the first segment are stored in a variable for later purposes. To visualize the dimensional scaling, two timings are separately captured, the time for linearization $t_\mathrm{Linearization}$ and the time required to fill the complete PS with the first segment $t_\mathrm{Polytope}$ for each reflection~$l$.

After linearizing and filling the PS, the subsequent intersection step is carried out to find the solution region; this step represents the most time-consuming part of the structure determination process, measured by $t_\mathrm{Intersection}$. At the end, the routine creates an HDF file and writes the information about the processed reflections $l$, the first segments, found solutions, the error on each solution, and the volume of each solution within the timing $t_\mathrm{Writing}$. In the case of MC simulations, we additionally write the information about the generated artificial atomic coordinates and the exact solution region that encloses the given structure. The total time $t_\mathrm{total}$ for the structure determination processes includes all four contributions separately for an increasing number of considered reflections~$l$. The individual times of the MC simulation in $\P{2}$ and $\P{3}$ are analyzed in detail in the Supplemental Material (Sec.~\numrange{3}{6}); below we only give $t_\mathrm{total}$.

Along with $t_\mathrm{total}$, the resulting average number of solutions and maximum error on $\bm{z}$ as defined in Eq.~\eqref{eq:error} are presented using the box plot analysis, see Sec.~\num{1} in SM for more details. Here, the \textit{average error} on $\bm{z}$ is obtained from the error on individual components, \ie $avg(\Delta z_1, \Delta z_2,\dots, \Delta z_m)$, and subsequent averaging over all MC instances. In addition, the area/volume of the solution region is given by the color bar. For the purpose of readability, we display the timings for adding two consecutive reflections~$l$. 

\subsubsection{Timing benchmark for 2D MC simulation}\label{sec:2DMC-time}

The MC results show the performance of the developed code and algorithm with respect to an increasing number of considered reflections. The individual timings are analyzed for the MC simulation in $\P{2}$ and presented in the Supplemental Material (Sec.~\numrange{3}{4}) for both intensity and amplitude approaches within EPA and non-EPA framework. 

\begin{figure}
    \centering
    \includegraphics[width=1\textwidth]{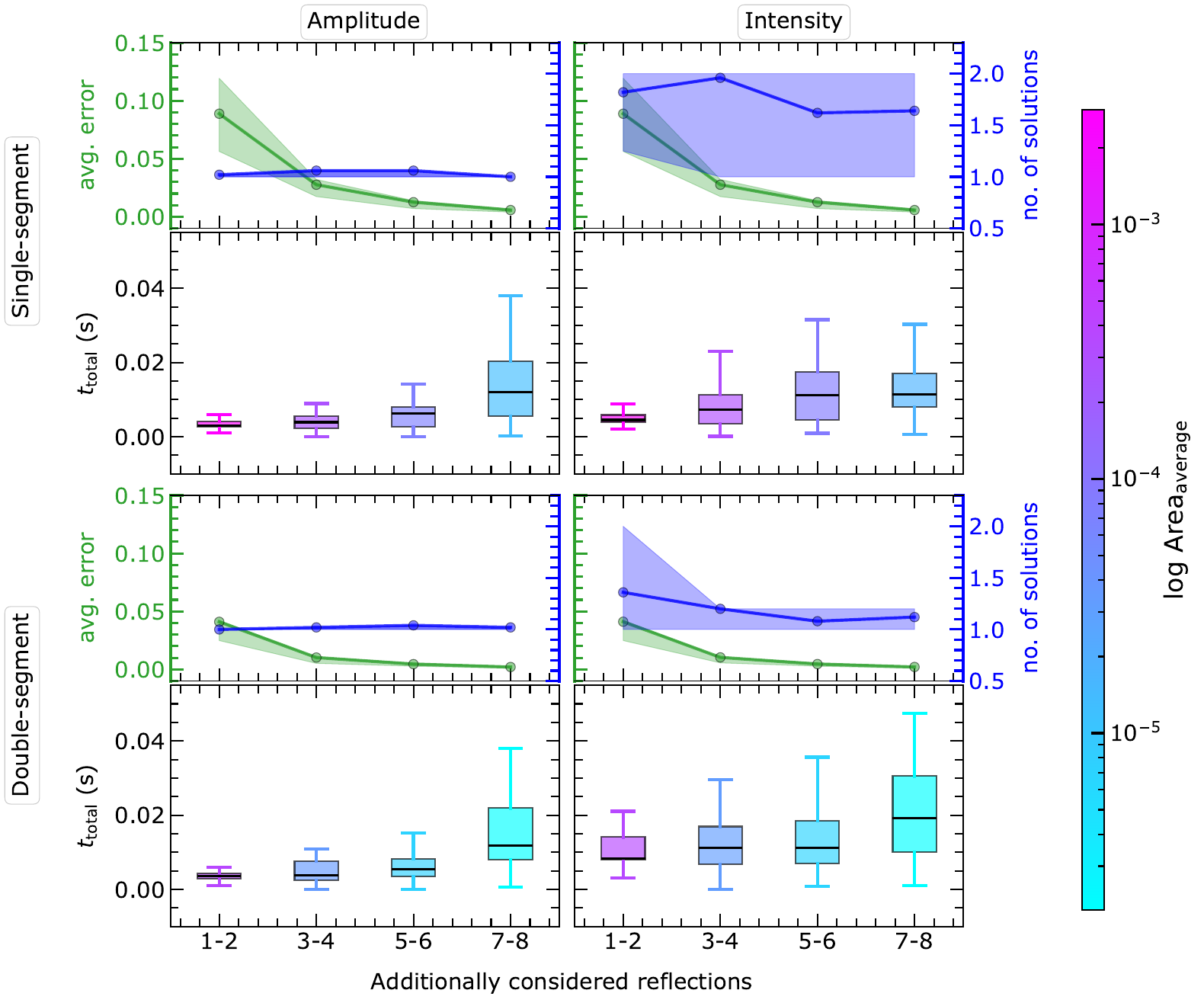}
    \caption{Incremental increases of $t_\mathrm{total}$ consumed for MC simulation within EPA framework in~$\P{2}$, including time for linearization, polytope creation, repeating in complete PS, intersection between successive reflections, and writing found solution details in an HDF file. The timing information is presented as boxplots, highlighting the median (black, horizontal line), quartiles (ends of the color box), and whiskers (here: complete data range), for details see Sec.~1 in SM. In addition, the average number of solutions and average maximum possible error on computed $\bm{z}$~coordinates are also summarized. These two characteristics are described by the median (circle) and the whisker position (envelope). The color of the boxes represents the calculated average area of the solution regions, which can be compared to the color code given in Fig.~\ref{fgr:2DMC_EPA}. The time information for the individual process is analyzed in detail in the SM, see Sec.~\num{3.1}~and~\num{3.2} in SM.}
    \label{fgr:2D_Totalt}
\end{figure}

\begin{figure}
    \centering
    \includegraphics[width=1\textwidth]{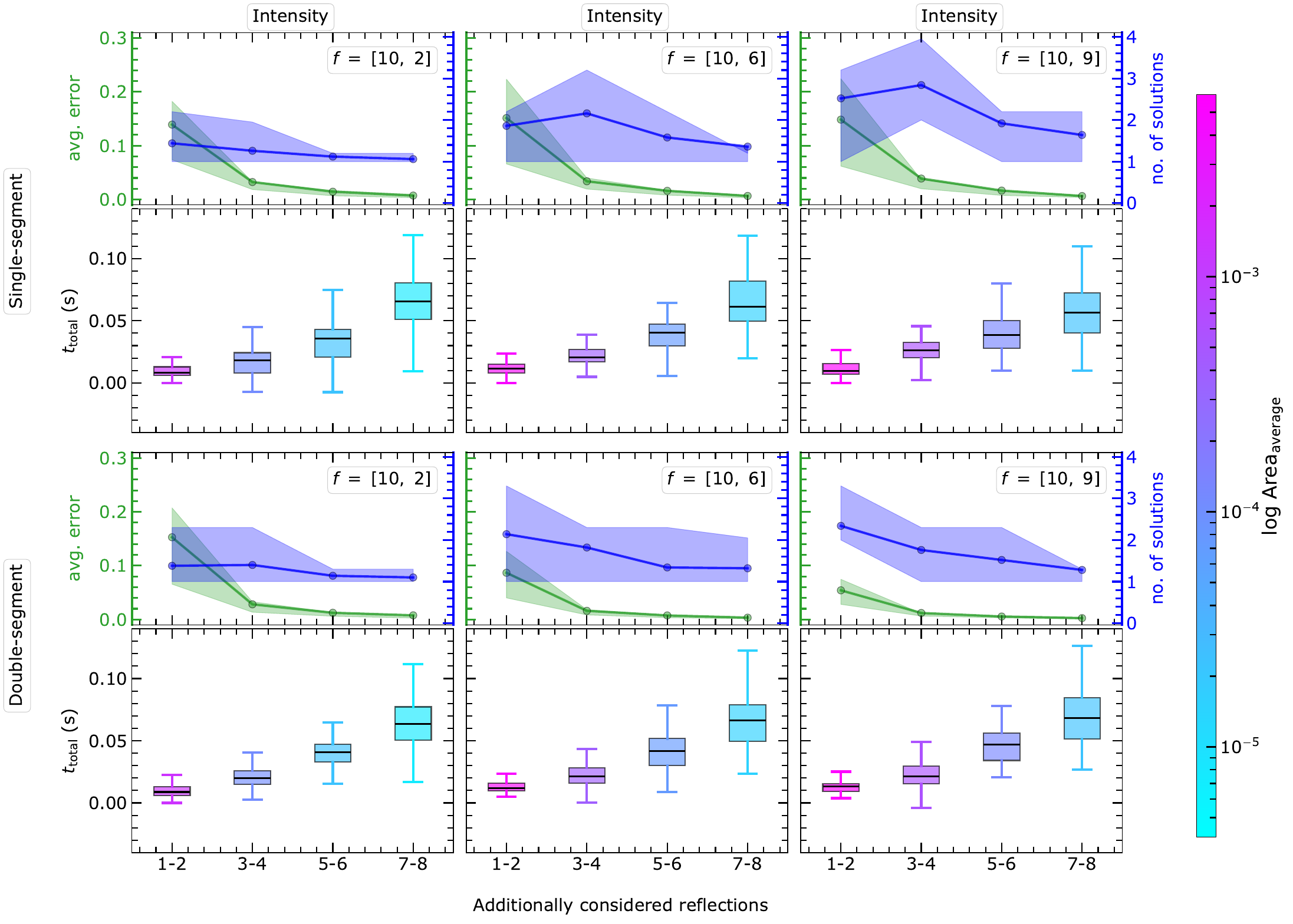}
    \caption{Incremental increases of $t_\mathrm{total}$ consumed for MC simulation within non-EPA framework in~$\P{2}$. The color of the boxes represents the calculated average area of solution regions which can be compared to the color code given in Fig.~\ref{fgr:2DMC_nEPA}. For further explanation see the caption in Fig.~\ref{fgr:2D_Totalt}. The time information for the individual process is analyzed in detail in the SM, see Sec.~\num{4.1}~and~\num{4.2} in SM.}
    \label{fgr:2D_TotaltnEPA}
\end{figure}

As presented in Fig.~\ref{fgr:2D_Totalt} and \ref{fgr:2D_TotaltnEPA} for EPA and non-EPA frameworks, a maximum of \SI{12}{\milli\s} and \SI{200}{\milli\s}, respectively, is spent to solve a structure with eight reflections irrespective of $f_i$ combinations. These timing observations are the result of executing the PSC code in the serial configuration. By parallelization of PSC routines, particularly the routine to fill the complete PS with segments, the time consumption can be severely reduced. Hence, the parallelization can be easily implemented in the future. 

Further, the time taken for each step in the code reveals that the two parts of completing the segments in PS and intersection making dominate, \cf Sec.~\num{3} and~\num{4} in SM. The repetition of the segment (Sec.~\ref{Completing the linearization}) can also be parallelized, which is a future task and has not yet been implemented in the code. The computational time consumption is expected to increase following the same trends in higher dimensions.

\subsubsection{Timing benchmark for 3D MC simulation}\label{3DMC-time}

\begin{figure}
    \centering
    \includegraphics[width=1\textwidth]{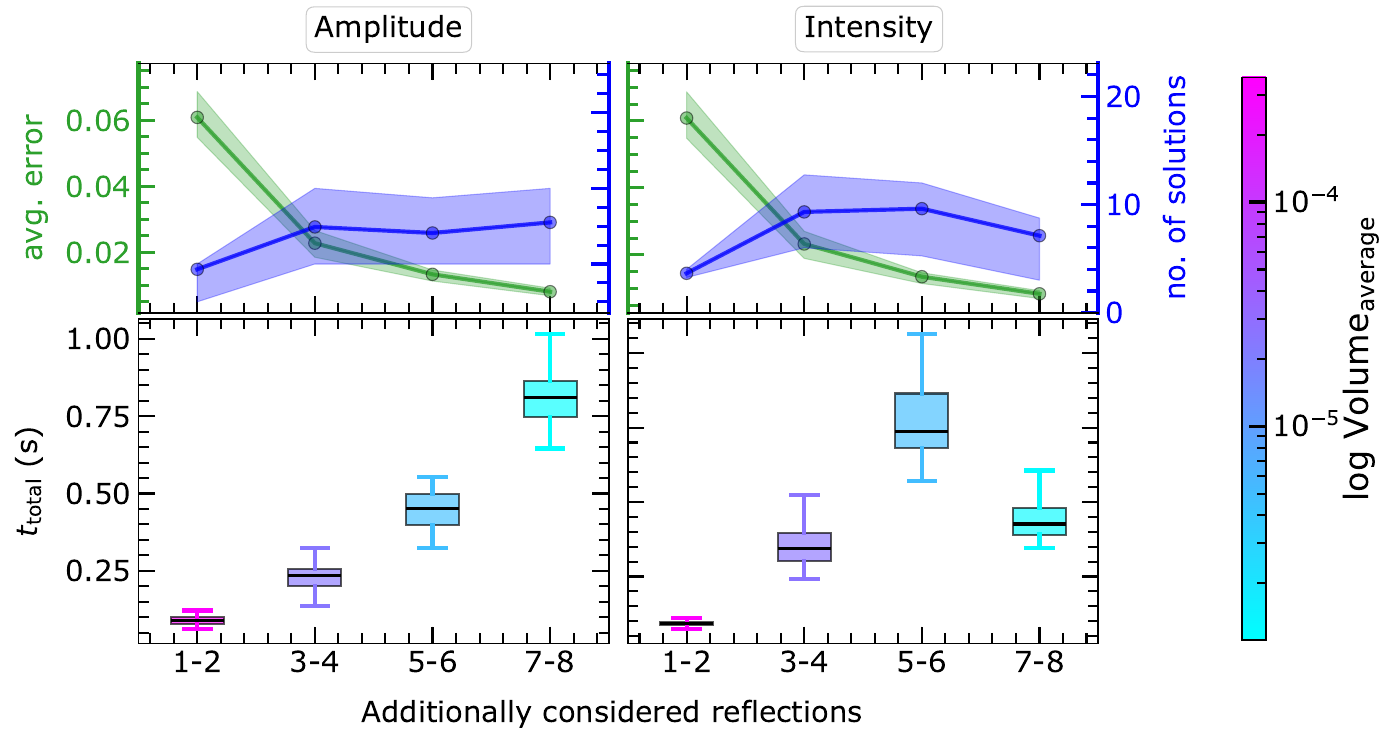}
    \caption{Incremental increases of $t_\mathrm{total}$ consumed in PSC simulation within EPA framework in~$\P{3}$, including linearization, polytopes creation and repeating in complete PS, intersection between successive reflections, and writing found solution details in a hierarchical data format file. The time information for the individual process is analyzed in detail in the SM, see Sec.~\num{5} in SM.}
    \label{fgr:3DMC_time_EPA}
\end{figure}

As already observed for the 2D cases, the repetition of segments through symmetry application and the intersection of polytopes consume more time than the linear approximation and the writing of solution details (see Sec.~\num{5} and \num{6} in SM). The total time of the entire process is given in Fig.~\ref{fgr:3DMC_time_EPA} for the EPA framework. The structure solution within the EPA model using eight reflections takes a maximum of \SI{10}{\s} and \SI{2}{\s} for the intensity and amplitude approach, respectively. The intensity approach is slower as the number of polytopes is significantly larger than for the amplitude approach.

\begin{figure}
    \centering
    \includegraphics[width=1\textwidth]{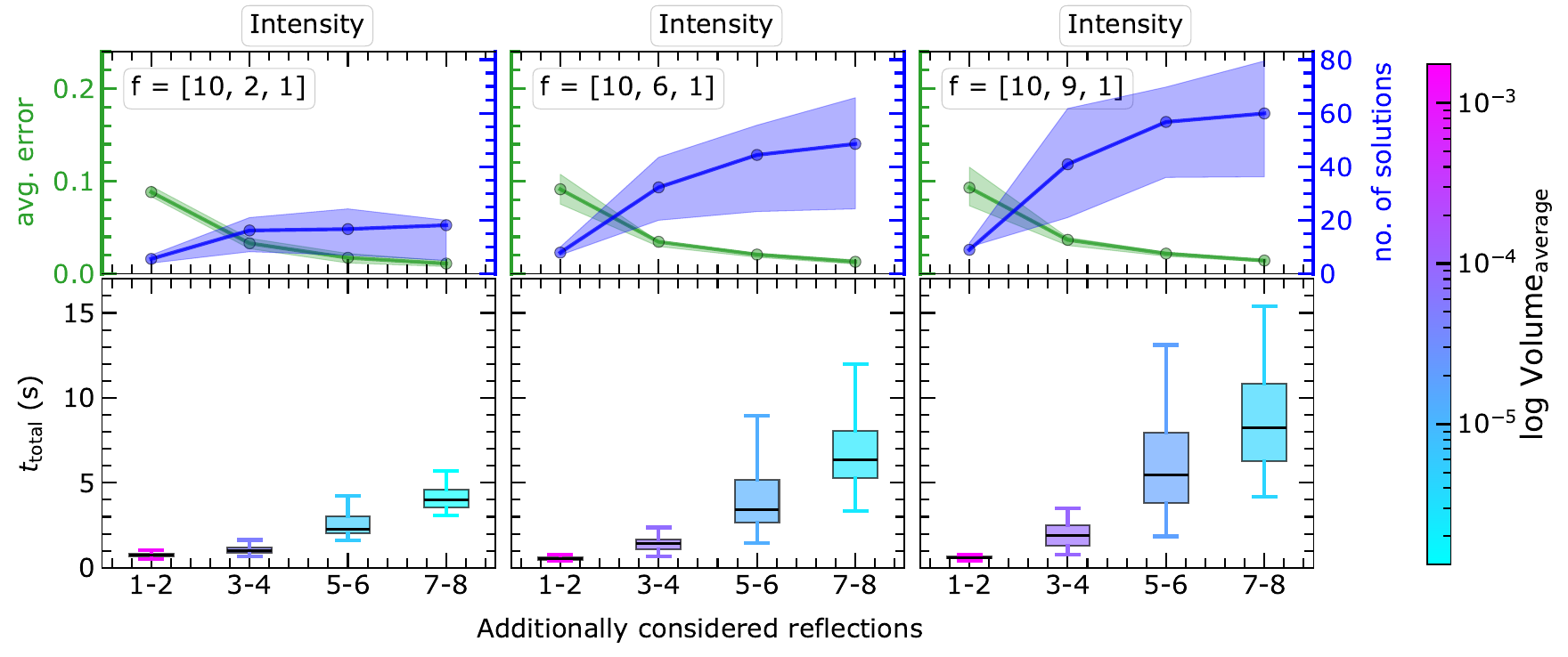}
    \caption{Incremental increases of $t_\mathrm{total}$ consumed in PSC simulation under the non-EPA framework in $\P{3}$, including linearization, polytopes creation and repeating in complete PS, intersection between successive reflections, and writing found solution details in a hierarchical data format file. The time information for the individual process is analyzed in detail in the SM, see Sec.~\num{6} in SM.}
    \label{fgr:3DMC_time_nEPA}
\end{figure}

In the case of non-EPA framework, \cf Fig.~\ref{fgr:3DMC_time_nEPA}, $t_\mathrm{total}$ strongly varies in dependence on the individual structure to be solved, which is visible from the outliers of the box plots. Due to the open topology, the number of possible solutions generally increases, and hence, $t_\mathrm{total}$ increases. Also, the computational workload increases continuously from the heavy-light-light to the heavy-heavy-light configuration. 

In certain cases, the number of solutions may decrease when adding more reflections in the calculation, which may reduce the computational load. Again, for each considered structure, we observed all possible solutions in one go, \cf the orange spheres in Fig.~\ref{fgr:3DMC_nEPA}. Hence, the PSC routines are robust for the different $f_i$ combinations in $\P{3}$. 
\clearpage


\section{Conclusion}

Thanks to the advancements in computational resources, we are now able to apply and implement the Parameter Space Concept proposed and developed by Fischer \textit{et al.} within the past 15 years. In the presented work, PSC has been enhanced to handle a broad spectrum of combinations of atomic scattering factors, making it suitable for realistic \xray data analysis. In this study, a concrete workflow is developed to initialize the obtained experimental/theoretical data, linearize the amplitude or intensity, span the PS with polytopes, carry out the intersection process, and perform the solution-finding routine. A stable algorithm has been developed to linearize the amplitudes and intensities under EPA and non-EPA schemes. It is observed that the developed algorithm can handle the open topologies in isosurfaces for all general cases. So far, the developed program effectively handles the structures in $\P{2}$ and $\P{3}$.

The linearization of the isosurfaces starts with defining the inner boundary using either the intersecting points of the isosurface with each axis for closed topologies or the period for open topologies. These intersecting points are utilized by invoking the singular value decomposition method to find the required normal vector~$\bm{\hat{n}}$. Then the respective tangent points on the isosurface are found by solving the parallel condition $\nabla{ \Hat{\mathcal{F}}}\cdot \bm{ \hat{n}~=~\pm 1}$  numerically using the least-squares method. The found tangent points are used to calculate the distance between the inner and outer boundaries to the origin. This implementation facilitates the generalization of PSC as well as the computational scaling of this step for the $m$-dimensional PS. In the next step, the required information (normal vector and boundary distances) can be converted into polytopes using efficient third-party Python libraries. The intersection process is also generalized to handle polytopes of any dimension. This step is carried out sequentially between successive reflections to scrutinize the feasible solution space for the given intensity or amplitude information.

The implemented PSC routines have been consecutively tested by employing Monte Carlo simulations. Artificial atomic structures have been generated randomly and treated with EPA and non-EPA models. A comprehensive analysis is conducted in both $\P{2}$ and $\P{3}$, allowing not only for a visual exploration of the combined effects of different atomic scattering factors, but also for qualitative analysis by evaluating the average values of area/volume of the polytopes and errors on the computed structures for each simulation step. A key observation from the simulations is that the curvature of the isosurface is predominantly influenced by heavy-light atom combinations and reflections that raise high structure amplitudes.

The presented results depict that with a limited number of reflections, a solution space with volume as small as \num{e-6} and extension on each structural degree of freedom $z_i$ in the order of \num{e-4} can be computed. A further advantage of PSC is that all possible structure solutions in accordance with the diffraction data appear in the PSC-determined solution volume.

In summary, the derived linearization-based PSC approach presents the means to solve crystal structures with equivalent and non-equivalent scattering factors with up to $m = 3$ degrees of freedom in the complete PS, to handle experimental diffraction data, and to explore all possible solutions in a single analysis. Additionally, the integration of third-party Python libraries for further data post-processing introduces new strategies for determining unknown structure parameters, offering an alternative to conventional FI refinement. The here presented routines as well as informative examples are publicly available on GitHub~\cite{code}.

\section{Outlook}

For the generalization of the PSC routines to an $m$-dimensional PS, efficient data handling becomes crucial and calls for further development of the current methodologies. It is worth mentioning that the current centrosymmetric constraint is expected to be resolved in the upcoming phase of PSC development, indicating a further advancement in PSC code applicability. 

Overall, PSC opens up new possibilities for improving the resolution and accuracy of atomic structure solutions employing resonant contrast~\cite{Zschornak2023}. The comparison between single- and double-segment linearization depicts that precise linearization would reduce the error from linearization approximation at the cost of computational effort and the number of polytopes to handle. Therefore, further development of the linearization schemes is inevitable.

In addition to the aforementioned challenges, the technical implementation of set theory operation critically shares a major computation time. The algorithm for the intersection of polytopes is kept in step-wise operation, recursively reducing the PS with each additional reflection~$l$. The themes for the next improvements are a sound parallel algorithm and innovative strategies to save the user time in handling the data. The sequence of processing the reflections may provide options to achieve the indented accuracy on atomic coordinates faster, and finding such a processing scheme is under development. Further enhancements will increase the possible degrees of freedom and aid in the capabilities to handle non-centrosymmetric structures.

Moreover, predicting the exact structure from multiple possible solutions is an open issue. A significant number of non-unique solutions will disappear once the sign of intensity is known and utilized. More may be identified as non-valid with sufficiently small intensity errors or the use of resonant contrast. The remaining non-unique solutions may be further analyzed and refined with modern state-of-the-art theoretical simulations like density functional theory~\cite{Blochl1994, Kresse1999,pbe} calculations, where the ground state energies of the predicted structures can be assessed to discriminate chemically unstable solutions.

In the present implementation, all routines are in principle generalized in such a way that they can be extended to handle arbitrary non-centrosymmetric structures in $m$-dimensional PS. Performance and reliability tests for $m \ge 4$ are currently in progress and will be the scope of further work.

\begin{acknowledgments}
This article is dedicated to the enduring legacy of the respected Prof. Dr. Karl Fischer (deceased in 2023), whose profound contributions to crystallography continue to inspire us. He was an invaluable part of the PSC project and we shall remember the valuable lessons learned under his insightful guidance and unwavering passion for knowledge. Prof. Fischer actively participated in weekly video conferences from his retirement home until shortly before the end of his life. His intellectual freshness, professional and expressly human advice and constant joy in exchanging ideas, even with young representatives of the next generation, was highly enriching for this work. The authors MZ and DCM gained professional and personal contact through the work of their common teacher, Prof. Peter Paufler (TU Dresden) and are very grateful for his continued support. MV, DM and MZ acknowledge funding by the DFG within the project DFG 442646446, ZS 120/5-1. MV thanks the Department of Information Service and Computing at Helmholtz-Zentrum Dresden-Rossendorf and the Center for Information Services and High-Performance Computing (ZIH), Technical University Dresden for providing extensive computing facilities. MV acknowledges and thanks Prof. Dr. Ralf Hielscher (Faculty of Mathematics and Informatics, TU Bergakademie Freiberg, Germany) for the fruitful discussion on the linearization of manifolds.
\end{acknowledgments}

\bibliography{apssamp}

\begin{thebibliography}{34}%
\makeatletter
\providecommand \@ifxundefined [1]{%
 \@ifx{#1\undefined}
}%
\providecommand \@ifnum [1]{%
 \ifnum #1\expandafter \@firstoftwo
 \else \expandafter \@secondoftwo
 \fi
}%
\providecommand \@ifx [1]{%
 \ifx #1\expandafter \@firstoftwo
 \else \expandafter \@secondoftwo
 \fi
}%
\providecommand \natexlab [1]{#1}%
\providecommand \enquote  [1]{``#1''}%
\providecommand \bibnamefont  [1]{#1}%
\providecommand \bibfnamefont [1]{#1}%
\providecommand \citenamefont [1]{#1}%
\providecommand \href@noop [0]{\@secondoftwo}%
\providecommand \href [0]{\begingroup \@sanitize@url \@href}%
\providecommand \@href[1]{\@@startlink{#1}\@@href}%
\providecommand \@@href[1]{\endgroup#1\@@endlink}%
\providecommand \@sanitize@url [0]{\catcode `\\12\catcode `\$12\catcode
  `\&12\catcode `\#12\catcode `\^12\catcode `\_12\catcode `\%12\relax}%
\providecommand \@@startlink[1]{}%
\providecommand \@@endlink[0]{}%
\providecommand \url  [0]{\begingroup\@sanitize@url \@url }%
\providecommand \@url [1]{\endgroup\@href {#1}{\urlprefix }}%
\providecommand \urlprefix  [0]{URL }%
\providecommand \Eprint [0]{\href }%
\providecommand \doibase [0]{https://doi.org/}%
\providecommand \selectlanguage [0]{\@gobble}%
\providecommand \bibinfo  [0]{\@secondoftwo}%
\providecommand \bibfield  [0]{\@secondoftwo}%
\providecommand \translation [1]{[#1]}%
\providecommand \BibitemOpen [0]{}%
\providecommand \bibitemStop [0]{}%
\providecommand \bibitemNoStop [0]{.\EOS\space}%
\providecommand \EOS [0]{\spacefactor3000\relax}%
\providecommand \BibitemShut  [1]{\csname bibitem#1\endcsname}%
\let\auto@bib@innerbib\@empty
\bibitem [{\citenamefont {Oszl{\'{a}}nyi}\ and\ \citenamefont
  {S{\"{u}}to}(2004)}]{Oszlanyi2004}%
  \BibitemOpen
  \bibfield  {author} {\bibinfo {author} {\bibfnamefont {G.}~\bibnamefont
  {Oszl{\'{a}}nyi}}\ and\ \bibinfo {author} {\bibfnamefont {A.}~\bibnamefont
  {S{\"{u}}to}},\ }\bibfield  {title} {\bibinfo {title} {{\textit{Ab initio}
  structure solution by charge flipping}},\ }\href
  {https://doi.org/10.1107/S0108767303027569} {\bibfield  {journal} {\bibinfo
  {journal} {Acta Cryst.~A}\ }\textbf {\bibinfo {volume} {60}},\ \bibinfo
  {pages} {134} (\bibinfo {year} {2004})}\BibitemShut {NoStop}%
\bibitem [{\citenamefont {Rothbauer}(1994)}]{Rothbauer1994}%
  \BibitemOpen
  \bibfield  {author} {\bibinfo {author} {\bibfnamefont {R.}~\bibnamefont
  {Rothbauer}},\ }\bibfield  {title} {\bibinfo {title} {{The phase equations of
  crystal structure analysis}},\ }\href
  {https://doi.org/0.1524/zkri.1994.209.7.578} {\bibfield  {journal} {\bibinfo
  {journal} {Z.~Kristallogr.}\ }\textbf {\bibinfo {volume} {209}},\ \bibinfo
  {pages} {578} (\bibinfo {year} {1994})}\BibitemShut {NoStop}%
\bibitem [{\citenamefont {Rothbauer}(1995)}]{Rothbauer1995}%
  \BibitemOpen
  \bibfield  {author} {\bibinfo {author} {\bibfnamefont {R.}~\bibnamefont
  {Rothbauer}},\ }\bibfield  {title} {\bibinfo {title} {{Crystal structure
  analysis by constrained optimisation}},\ }\href
  {https://doi.org/10.1524/zkri.1995.210.4.255} {\bibfield  {journal} {\bibinfo
   {journal} {Z.~Kristallogr.}\ }\textbf {\bibinfo {volume} {210}},\ \bibinfo
  {pages} {255} (\bibinfo {year} {1995})}\BibitemShut {NoStop}%
\bibitem [{\citenamefont {Rothbauer}(1998)}]{Rothbauer1998}%
  \BibitemOpen
  \bibfield  {author} {\bibinfo {author} {\bibfnamefont {R.}~\bibnamefont
  {Rothbauer}},\ }\bibfield  {title} {\bibinfo {title} {{The functional
  dependence of the coordinates of atoms on form and structure factors}},\
  }\href {https://doi.org/10.1524/zkri.1998.213.4.195} {\bibfield  {journal}
  {\bibinfo  {journal} {Z.~Kristallogr.}\ }\textbf {\bibinfo {volume} {213}},\
  \bibinfo {pages} {195} (\bibinfo {year} {1998})}\BibitemShut {NoStop}%
\bibitem [{\citenamefont {Navaza}\ and\ \citenamefont
  {Silva}(1979)}]{Navaza1979}%
  \BibitemOpen
  \bibfield  {author} {\bibinfo {author} {\bibfnamefont {J.}~\bibnamefont
  {Navaza}}\ and\ \bibinfo {author} {\bibfnamefont {A.~M.}\ \bibnamefont
  {Silva}},\ }\bibfield  {title} {\bibinfo {title} {{A geometrical-approach to
  solving crystal structures}},\ }\href
  {https://doi.org/10.1107/S0567739479000577} {\bibfield  {journal} {\bibinfo
  {journal} {Acta Cryst.}\ }\textbf {\bibinfo {volume} {A35}},\ \bibinfo
  {pages} {266} (\bibinfo {year} {1979})}\BibitemShut {NoStop}%
\bibitem [{\citenamefont {Cervellino}\ and\ \citenamefont
  {Ciccariello}(2005)}]{Cervellino2005}%
  \BibitemOpen
  \bibfield  {author} {\bibinfo {author} {\bibfnamefont {A.}~\bibnamefont
  {Cervellino}}\ and\ \bibinfo {author} {\bibfnamefont {S.}~\bibnamefont
  {Ciccariello}},\ }\bibfield  {title} {\bibinfo {title} {{The algebraic
  approach to the phase problem}},\ }\href
  {https://doi.org/10.1107/S0108767305019860} {\bibfield  {journal} {\bibinfo
  {journal} {Acta Cryst.~A}\ }\textbf {\bibinfo {volume} {61}},\ \bibinfo
  {pages} {494} (\bibinfo {year} {2005})}\BibitemShut {NoStop}%
\bibitem [{\citenamefont {Toby}(2024)}]{Tobyactacrys}%
  \BibitemOpen
  \bibfield  {author} {\bibinfo {author} {\bibfnamefont {B.~H.}\ \bibnamefont
  {Toby}},\ }\bibfield  {title} {\bibinfo {title} {{A simple solution to the
  Rietveld refinement recipe problem}},\ }\href
  {https://doi.org/10.1107/S1600576723011032} {\bibfield  {journal} {\bibinfo
  {journal} {J.~Appl. Cryst.}\ }\textbf {\bibinfo {volume} {57}},\ \bibinfo
  {pages} {175} (\bibinfo {year} {2024})}\BibitemShut {NoStop}%
\bibitem [{\citenamefont {Shi}(2022)}]{shiactacrys}%
  \BibitemOpen
  \bibfield  {author} {\bibinfo {author} {\bibfnamefont {H.}~\bibnamefont
  {Shi}},\ }\bibfield  {title} {\bibinfo {title} {{Determining lattice
  parameters from two electron diffraction patterns}},\ }\href
  {https://doi.org/10.1107/S1600576722004630} {\bibfield  {journal} {\bibinfo
  {journal} {J.~Appl. Cryst.}\ }\textbf {\bibinfo {volume} {55}},\ \bibinfo
  {pages} {669} (\bibinfo {year} {2022})}\BibitemShut {NoStop}%
\bibitem [{\citenamefont {Munteanu}\ \emph {et~al.}(2024)\citenamefont
  {Munteanu}, \citenamefont {Starostin}, \citenamefont {Greco}, \citenamefont
  {Pithan}, \citenamefont {Gerlach}, \citenamefont {Hinderhofer}, \citenamefont
  {Kowarik},\ and\ \citenamefont {Schreiber}}]{Munteanuactacrys}%
  \BibitemOpen
  \bibfield  {author} {\bibinfo {author} {\bibfnamefont {V.}~\bibnamefont
  {Munteanu}}, \bibinfo {author} {\bibfnamefont {V.}~\bibnamefont {Starostin}},
  \bibinfo {author} {\bibfnamefont {A.}~\bibnamefont {Greco}}, \bibinfo
  {author} {\bibfnamefont {L.}~\bibnamefont {Pithan}}, \bibinfo {author}
  {\bibfnamefont {A.}~\bibnamefont {Gerlach}}, \bibinfo {author} {\bibfnamefont
  {A.}~\bibnamefont {Hinderhofer}}, \bibinfo {author} {\bibfnamefont
  {S.}~\bibnamefont {Kowarik}},\ and\ \bibinfo {author} {\bibfnamefont
  {F.}~\bibnamefont {Schreiber}},\ }\bibfield  {title} {\bibinfo {title}
  {{Neural network analysis of neutron and X-ray reflectivity data
  incorporating prior knowledge}},\ }\href
  {https://doi.org/10.1107/S1600576724002115} {\bibfield  {journal} {\bibinfo
  {journal} {J.~Appl. Cryst.}\ }\textbf {\bibinfo {volume} {57}},\ \bibinfo
  {pages} {456} (\bibinfo {year} {2024})}\BibitemShut {NoStop}%
\bibitem [{\citenamefont {Billinge}\ and\ \citenamefont
  {Proffen}(2024)}]{Billing}%
  \BibitemOpen
  \bibfield  {author} {\bibinfo {author} {\bibfnamefont {S.~J.~L.}\
  \bibnamefont {Billinge}}\ and\ \bibinfo {author} {\bibfnamefont
  {T.}~\bibnamefont {Proffen}},\ }\bibfield  {title} {\bibinfo {title}
  {{Machine learning in crystallography and structural science}},\ }\href
  {https://doi.org/10.1107/S2053273324000172} {\bibfield  {journal} {\bibinfo
  {journal} {Acta Cryst.~A}\ }\textbf {\bibinfo {volume} {80}},\ \bibinfo
  {pages} {139} (\bibinfo {year} {2024})}\BibitemShut {NoStop}%
\bibitem [{\citenamefont {Harrison}(1993)}]{Harrison1993}%
  \BibitemOpen
  \bibfield  {author} {\bibinfo {author} {\bibfnamefont {R.~W.}\ \bibnamefont
  {Harrison}},\ }\bibfield  {title} {\bibinfo {title} {{Phase problem in
  crystallography}},\ }\href {https://doi.org/10.1364/JOSAA.10.001046}
  {\bibfield  {journal} {\bibinfo  {journal} {J.~Opt. Soc. Amer.}\ }\textbf
  {\bibinfo {volume} {10}},\ \bibinfo {pages} {1046} (\bibinfo {year}
  {1993})}\BibitemShut {NoStop}%
\bibitem [{\citenamefont {Fischer}\ \emph {et~al.}(2005)\citenamefont
  {Fischer}, \citenamefont {Kirfel},\ and\ \citenamefont
  {Zimmermann}}]{Fischer2005}%
  \BibitemOpen
  \bibfield  {author} {\bibinfo {author} {\bibfnamefont {K.~F.}\ \bibnamefont
  {Fischer}}, \bibinfo {author} {\bibfnamefont {A.}~\bibnamefont {Kirfel}},\
  and\ \bibinfo {author} {\bibfnamefont {H.~W.}\ \bibnamefont {Zimmermann}},\
  }\bibfield  {title} {\bibinfo {title} {{Structure determination without
  Fourier inversion. Part I. Unique results for centrosymmetric examples}},\
  }\href {https://doi.org/10.1524/zkri.220.7.643.67099} {\bibfield  {journal}
  {\bibinfo  {journal} {Z.~Kristallogr.}\ }\textbf {\bibinfo {volume} {220}},\
  \bibinfo {pages} {643} (\bibinfo {year} {2005})}\BibitemShut {NoStop}%
\bibitem [{\citenamefont {Fischer}\ \emph {et~al.}(2008)\citenamefont
  {Fischer}, \citenamefont {Kirfel},\ and\ \citenamefont
  {Zimmermann}}]{Fischer2008}%
  \BibitemOpen
  \bibfield  {author} {\bibinfo {author} {\bibfnamefont {K.~F.}\ \bibnamefont
  {Fischer}}, \bibinfo {author} {\bibfnamefont {A.}~\bibnamefont {Kirfel}},\
  and\ \bibinfo {author} {\bibfnamefont {H.}~\bibnamefont {Zimmermann}},\
  }\bibfield  {title} {\bibinfo {title} {{A Concept for Crystal Structure
  Determination without Fourier Inversion: Some Steps towards Application}},\
  }\href {https://hrcak.srce.hr/28508} {\bibfield  {journal} {\bibinfo
  {journal} {Croat. Chem. Acta}\ }\textbf {\bibinfo {volume} {81}},\ \bibinfo
  {pages} {381} (\bibinfo {year} {2008})}\BibitemShut {NoStop}%
\bibitem [{\citenamefont {Kirfel}\ and\ \citenamefont
  {Fischer}(2009)}]{Fischer2009}%
  \BibitemOpen
  \bibfield  {author} {\bibinfo {author} {\bibfnamefont {A.}~\bibnamefont
  {Kirfel}}\ and\ \bibinfo {author} {\bibfnamefont {K.~F.}\ \bibnamefont
  {Fischer}},\ }\bibfield  {title} {\bibinfo {title} {{Structure determination
  without Fourier inversion. Part IV: Using quasi-normalized data}},\ }\href
  {https://doi.org/10.1524/zkri.2009.1130} {\bibfield  {journal} {\bibinfo
  {journal} {Z.~Kristallogr.}\ }\textbf {\bibinfo {volume} {224}},\ \bibinfo
  {pages} {325} (\bibinfo {year} {2009})}\BibitemShut {NoStop}%
\bibitem [{\citenamefont {Zschornak}\ \emph {et~al.}(2024)\citenamefont
  {Zschornak}, \citenamefont {Wagner}, \citenamefont {Nentwich}, \citenamefont
  {Vallinayagam},\ and\ \citenamefont {Fischer}}]{Zschornak2023}%
  \BibitemOpen
  \bibfield  {author} {\bibinfo {author} {\bibfnamefont {M.}~\bibnamefont
  {Zschornak}}, \bibinfo {author} {\bibfnamefont {C.}~\bibnamefont {Wagner}},
  \bibinfo {author} {\bibfnamefont {M.}~\bibnamefont {Nentwich}}, \bibinfo
  {author} {\bibfnamefont {M.}~\bibnamefont {Vallinayagam}},\ and\ \bibinfo
  {author} {\bibfnamefont {K.~F.}\ \bibnamefont {Fischer}},\ }\bibfield
  {title} {\bibinfo {title} {Advances in the parameter space concept towards
  picometer precise crystal structure refinement -- a resolution study},\
  }\href {https://doi.org/10.3390/cryst14080684} {\bibfield  {journal}
  {\bibinfo  {journal} {Crystals}\ }\textbf {\bibinfo {volume} {14}},\ \bibinfo
  {pages} {684} (\bibinfo {year} {2024})}\BibitemShut {NoStop}%
\bibitem [{\citenamefont {Knop}(1989)}]{knop_thesis}%
  \BibitemOpen
  \bibfield  {author} {\bibinfo {author} {\bibfnamefont {W.}~\bibnamefont
  {Knop}},\ }\emph {\bibinfo {title} {{Analytische Methode zur
  Strukturbestimmung bei energiedispersiven Lauemessungen}}},\ \href@noop {}
  {\bibinfo {type} {dissertation}},\ \bibinfo  {school} {Universit{\"a}t
  Saarbr{\"u}cken} (\bibinfo {year} {1989})\BibitemShut {NoStop}%
\bibitem [{\citenamefont {Pilz}(1996)}]{pliz_thesis}%
  \BibitemOpen
  \bibfield  {author} {\bibinfo {author} {\bibfnamefont {K.}~\bibnamefont
  {Pilz}},\ }\emph {\bibinfo {title} {{Weiterentwicklung und Anwendung einer
  algebraischen Methode zur Teilstrukturbestimmung, ein Beitrag zur
  Eindeutigkeit von Strukturanalysen}}},\ \href@noop {} {\bibinfo {type}
  {dissertation}},\ \bibinfo  {school} {Universit{\"a}t Saarbr{\"u}cken}
  (\bibinfo {year} {1996})\BibitemShut {NoStop}%
\bibitem [{\citenamefont {Ott}(1928)}]{Ott1927}%
  \BibitemOpen
  \bibfield  {author} {\bibinfo {author} {\bibfnamefont {H.}~\bibnamefont
  {Ott}},\ }\bibfield  {title} {\bibinfo {title} {{IX. Zur Methodik der
  Strukturanalyse}},\ }\href {https://doi.org/10.1524/zkri.1928.66.1.136}
  {\bibfield  {journal} {\bibinfo  {journal} {Z.~Kristallogr.}\ }\textbf
  {\bibinfo {volume} {66}},\ \bibinfo {pages} {136} (\bibinfo {year}
  {1928})}\BibitemShut {NoStop}%
\bibitem [{\citenamefont {Kirfel}\ \emph {et~al.}(2006)\citenamefont {Kirfel},
  \citenamefont {Fischer},\ and\ \citenamefont {Zimmermann}}]{Fischer2006}%
  \BibitemOpen
  \bibfield  {author} {\bibinfo {author} {\bibfnamefont {A.}~\bibnamefont
  {Kirfel}}, \bibinfo {author} {\bibfnamefont {K.~F.}\ \bibnamefont
  {Fischer}},\ and\ \bibinfo {author} {\bibfnamefont {H.~W.}\ \bibnamefont
  {Zimmermann}},\ }\bibfield  {title} {\bibinfo {title} {{Structure
  determination without Fourier inversion. Part II: The use of intensity ratios
  and inequalities}},\ }\href {https://doi.org/10.1524/zkri.2006.221.10.673}
  {\bibfield  {journal} {\bibinfo  {journal} {Z.~Kristallogr.}\ }\textbf
  {\bibinfo {volume} {221}},\ \bibinfo {pages} {673} (\bibinfo {year}
  {2006})}\BibitemShut {NoStop}%
\bibitem [{\citenamefont {Woolfson}\ and\ \citenamefont
  {Hai-Fu}(1995)}]{woolfson_hai-fu_1995}%
  \BibitemOpen
  \bibfield  {author} {\bibinfo {author} {\bibfnamefont {M.~M.}\ \bibnamefont
  {Woolfson}}\ and\ \bibinfo {author} {\bibfnamefont {F.}~\bibnamefont
  {Hai-Fu}},\ }\href {https://doi.org/10.1017/CBO9780511599866} {\emph
  {\bibinfo {title} {{Physical and Non-Physical Methods of Solving Crystal
  Structures}}}}\ (\bibinfo  {publisher} {Cambridge University Press},\
  \bibinfo {year} {1995})\BibitemShut {NoStop}%
\bibitem [{\citenamefont {Lozada-Cruz}(2020)}]{MVT1}%
  \BibitemOpen
  \bibfield  {author} {\bibinfo {author} {\bibfnamefont {G.}~\bibnamefont
  {Lozada-Cruz}},\ }\bibfield  {title} {\bibinfo {title} {Some variants of
  cauchy's mean value theorem},\ }\href
  {https://doi.org/10.1080/0020739X.2019.1703150} {\bibfield  {journal}
  {\bibinfo  {journal} {Int. J.~Math. Edu. Sci. Technol.}\ }\textbf {\bibinfo
  {volume} {51}},\ \bibinfo {pages} {1155} (\bibinfo {year}
  {2020})}\BibitemShut {NoStop}%
\bibitem [{\citenamefont {Sahoo}\ and\ \citenamefont {Riedel}(1998)}]{MVT2}%
  \BibitemOpen
  \bibfield  {author} {\bibinfo {author} {\bibfnamefont {P.~K.}\ \bibnamefont
  {Sahoo}}\ and\ \bibinfo {author} {\bibfnamefont {T.}~\bibnamefont {Riedel}},\
  }\href {https://doi.org/10.1142/3857} {\emph {\bibinfo {title} {Mean Value
  Theorems and Functional Equations}}}\ (\bibinfo  {publisher} {World
  Scientific},\ \bibinfo {year} {1998})\BibitemShut {NoStop}%
\bibitem [{\citenamefont {Group}\ and\ \citenamefont {Koziol}(2020)}]{HDF}%
  \BibitemOpen
  \bibfield  {author} {\bibinfo {author} {\bibfnamefont {T.~H.}\ \bibnamefont
  {Group}}\ and\ \bibinfo {author} {\bibfnamefont {Q.}~\bibnamefont {Koziol}},\
  }\href {https://doi.org/10.11578/dc.20180330.1} {\bibinfo {title}
  {Hdf5-version 1.12.0}} (\bibinfo {year} {2020})\BibitemShut {NoStop}%
\bibitem [{\citenamefont {Hobson}(1909)}]{Hobson1909}%
  \BibitemOpen
  \bibfield  {author} {\bibinfo {author} {\bibfnamefont {E.~W.}\ \bibnamefont
  {Hobson}},\ }\bibfield  {title} {\bibinfo {title} {{On the Second Mean-Value
  Theorem of the Integral Calculus}},\ }\href
  {https://doi.org/10.1112/plms/s2-7.1.14} {\bibfield  {journal} {\bibinfo
  {journal} {Z.~Kristallogr.}\ }\textbf {\bibinfo {volume} {s2.7}},\ \bibinfo
  {pages} {14} (\bibinfo {year} {1909})}\BibitemShut {NoStop}%
\bibitem [{\citenamefont {Dekking}\ \emph {et~al.}(2005)\citenamefont
  {Dekking}, \citenamefont {Kraaikamp}, \citenamefont {Lopuha{\"a}},\ and\
  \citenamefont {Meester}}]{dekking2005}%
  \BibitemOpen
  \bibfield  {author} {\bibinfo {author} {\bibfnamefont {F.~M.}\ \bibnamefont
  {Dekking}}, \bibinfo {author} {\bibfnamefont {C.}~\bibnamefont {Kraaikamp}},
  \bibinfo {author} {\bibfnamefont {H.~P.}\ \bibnamefont {Lopuha{\"a}}},\ and\
  \bibinfo {author} {\bibfnamefont {L.~E.}\ \bibnamefont {Meester}},\
  }\href@noop {} {\emph {\bibinfo {title} {A Modern Introduction to Probability
  and Statistics: Understanding Why and How}}},\ Springer Texts in Statistics\
  (\bibinfo  {publisher} {Springer},\ \bibinfo {address} {London},\ \bibinfo
  {year} {2005})\BibitemShut {NoStop}%
\bibitem [{\citenamefont {Lawson}\ and\ \citenamefont
  {Hanson}(1974)}]{lawson1974}%
  \BibitemOpen
  \bibfield  {author} {\bibinfo {author} {\bibfnamefont {C.~L.}\ \bibnamefont
  {Lawson}}\ and\ \bibinfo {author} {\bibfnamefont {R.~J.}\ \bibnamefont
  {Hanson}},\ }\href@noop {} {\emph {\bibinfo {title} {Solving Least Squares
  Problems}}}\ (\bibinfo  {publisher} {Prentice-Hall},\ \bibinfo {address}
  {Englewood Cliffs, NJ},\ \bibinfo {year} {1974})\ \bibinfo {note} {reprinted
  by SIAM, 1995}\BibitemShut {NoStop}%
\bibitem [{\citenamefont {Campbell}\ \emph {et~al.}(2021)\citenamefont
  {Campbell}, \citenamefont {Stokes}, \citenamefont {Averett}, \citenamefont
  {Machlus},\ and\ \citenamefont {Yost}}]{Campbellactacrys}%
  \BibitemOpen
  \bibfield  {author} {\bibinfo {author} {\bibfnamefont {B.~J.}\ \bibnamefont
  {Campbell}}, \bibinfo {author} {\bibfnamefont {H.~T.}\ \bibnamefont
  {Stokes}}, \bibinfo {author} {\bibfnamefont {T.~B.}\ \bibnamefont {Averett}},
  \bibinfo {author} {\bibfnamefont {S.}~\bibnamefont {Machlus}},\ and\ \bibinfo
  {author} {\bibfnamefont {C.~J.}\ \bibnamefont {Yost}},\ }\bibfield  {title}
  {\bibinfo {title} {{Theoretical and computational improvements to the
  algebraic method for discovering cooperative rigid-unit modes}},\ }\href
  {https://doi.org/10.1107/S1600576721009341} {\bibfield  {journal} {\bibinfo
  {journal} {J.~Appl. Cryst.}\ }\textbf {\bibinfo {volume} {54}},\ \bibinfo
  {pages} {1664} (\bibinfo {year} {2021})}\BibitemShut {NoStop}%
\bibitem [{\citenamefont {Filippidis}\ \emph {et~al.}(2016)\citenamefont
  {Filippidis}, \citenamefont {Dathathri}, \citenamefont {Livingston},
  \citenamefont {Ozay},\ and\ \citenamefont {Murray}}]{polytope}%
  \BibitemOpen
  \bibfield  {author} {\bibinfo {author} {\bibfnamefont {I.}~\bibnamefont
  {Filippidis}}, \bibinfo {author} {\bibfnamefont {S.}~\bibnamefont
  {Dathathri}}, \bibinfo {author} {\bibfnamefont {S.~C.}\ \bibnamefont
  {Livingston}}, \bibinfo {author} {\bibfnamefont {N.}~\bibnamefont {Ozay}},\
  and\ \bibinfo {author} {\bibfnamefont {R.~M.}\ \bibnamefont {Murray}},\
  }\bibfield  {title} {\bibinfo {title} {Control design for hybrid systems with
  tulip: The temporal logic planning toolbox},\ }in\ \href
  {https://doi.org/10.1109/CCA.2016.7587949} {\emph {\bibinfo {booktitle} {2016
  IEEE Conference on Control Applications (CCA)}}}\ (\bibinfo {year} {2016})\
  p.\ \bibinfo {pages} {1030}\BibitemShut {NoStop}%
\bibitem [{\citenamefont {Gillies}\ \emph {et~al.}(2022)\citenamefont {Gillies}
  \emph {et~al.}}]{shapely}%
  \BibitemOpen
  \bibfield  {author} {\bibinfo {author} {\bibfnamefont {S.}~\bibnamefont
  {Gillies}} \emph {et~al.},\ }\href {https://doi.org/10.5281/zenodo.10671398}
  {\bibinfo {title} {{Shapely User Manual}}} (\bibinfo {year}
  {2022})\BibitemShut {NoStop}%
\bibitem [{\citenamefont {Androsov}\ and\ \citenamefont
  {Shary}(2022)}]{intvalpy}%
  \BibitemOpen
  \bibfield  {author} {\bibinfo {author} {\bibfnamefont {A.~S.}\ \bibnamefont
  {Androsov}}\ and\ \bibinfo {author} {\bibfnamefont {S.~P.}\ \bibnamefont
  {Shary}},\ }\bibfield  {title} {\bibinfo {title} {Intvalpy -- a python
  interval computation library},\ }\href
  {https://doi.org/10.25205/1818-7900-2022-20-4-5-23} {\bibfield  {journal}
  {\bibinfo  {journal} {Vestnik NSU. Series: Information Technologies}\
  }\textbf {\bibinfo {volume} {20}},\ \bibinfo {pages} {5} (\bibinfo {year}
  {2022})}\BibitemShut {NoStop}%
\bibitem [{\citenamefont {Vallinayagam}\ \emph {et~al.}(2024)\citenamefont
  {Vallinayagam}, \citenamefont {Nentwich},\ and\ \citenamefont
  {Zschornak}}]{code}%
  \BibitemOpen
  \bibfield  {author} {\bibinfo {author} {\bibfnamefont {M.}~\bibnamefont
  {Vallinayagam}}, \bibinfo {author} {\bibfnamefont {M.}~\bibnamefont
  {Nentwich}},\ and\ \bibinfo {author} {\bibfnamefont {M.}~\bibnamefont
  {Zschornak}},\ }\href {https://github.com/mvnayagam/pypsc.git} {\bibinfo
  {title} {pypsc -- parameter space concept code}},\ \bibinfo {howpublished}
  {GitHub} (\bibinfo {year} {2024})\BibitemShut {NoStop}%
\bibitem [{\citenamefont {Bl{\"o}chl}(1994)}]{Blochl1994}%
  \BibitemOpen
  \bibfield  {author} {\bibinfo {author} {\bibfnamefont {P.~E.}\ \bibnamefont
  {Bl{\"o}chl}},\ }\bibfield  {title} {\bibinfo {title} {{Projector
  augmented-wave method}},\ }\href {https://doi.org/10.1103/PhysRevB.50.17953}
  {\bibfield  {journal} {\bibinfo  {journal} {Phys. Rev.~B}\ }\textbf {\bibinfo
  {volume} {50}},\ \bibinfo {pages} {17953} (\bibinfo {year}
  {1994})}\BibitemShut {NoStop}%
\bibitem [{\citenamefont {Kresse}\ and\ \citenamefont
  {Joubert}(1999)}]{Kresse1999}%
  \BibitemOpen
  \bibfield  {author} {\bibinfo {author} {\bibfnamefont {G.}~\bibnamefont
  {Kresse}}\ and\ \bibinfo {author} {\bibfnamefont {D.}~\bibnamefont
  {Joubert}},\ }\bibfield  {title} {\bibinfo {title} {{From ultrasoft
  pseudopotentials to the projector augmented-wave method}},\ }\href
  {https://doi.org/10.1103/PhysRevB.59.1758} {\bibfield  {journal} {\bibinfo
  {journal} {Phys. Rev.~B}\ }\textbf {\bibinfo {volume} {59}},\ \bibinfo
  {pages} {1758} (\bibinfo {year} {1999})}\BibitemShut {NoStop}%
\bibitem [{\citenamefont {Perdew}\ \emph {et~al.}(1996)\citenamefont {Perdew},
  \citenamefont {Burke},\ and\ \citenamefont {Ernzerhof}}]{pbe}%
  \BibitemOpen
  \bibfield  {author} {\bibinfo {author} {\bibfnamefont {J.~P.}\ \bibnamefont
  {Perdew}}, \bibinfo {author} {\bibfnamefont {K.}~\bibnamefont {Burke}},\ and\
  \bibinfo {author} {\bibfnamefont {M.}~\bibnamefont {Ernzerhof}},\ }\bibfield
  {title} {\bibinfo {title} {Generalized gradient approximation made simple},\
  }\href {https://doi.org/10.1103/PhysRevLett.77.3865} {\bibfield  {journal}
  {\bibinfo  {journal} {Phys. Rev. Lett.}\ }\textbf {\bibinfo {volume} {77}},\
  \bibinfo {pages} {3865} (\bibinfo {year} {1996})}\BibitemShut {NoStop}%
\end{thebibliography}%
\end{document}